\documentclass[11pt]{article}
\usepackage{epsf,latexsym,amsfonts}
\newcommand{\pr}{{Phys.\ Rev.\/}}
\newcommand{\np}{{Nucl.\ Phys.\/}}
\newcommand{\pl}{{Phys.\ Lett.\/}}

\title{Flux tubes and their interaction in $U(1)$ lattice gauge theory
\thanks{Supported by Fonds zur F\"orderung der wissenschaftlichen Forschung, Proj.~P11156-PHY}}
 
\author{Martin Zach, Manfried Faber and Peter Skala\\
Institut f\"ur Kernphysik, Technische Universit\"at Wien, \\ A-1040 Vienna, Austria}

\date{}

\begin{document}

\maketitle

\begin{abstract}
We investigate singly and doubly charged flux tubes in $U(1)$ lattice gauge theory. By simulating the dually transformed path integral we are able to consider large flux tube lengths, low temperatures, and multiply charged systems without loss of numerical precision. We simulate flux tubes between static sources as well as periodically closed flux tubes, calculating flux tube profiles, the total field energy and the free energy. Our main results are that the string tension in both three and four dimensions scales proportionally to the charge -- which is in contrast to previous lattice results -- and that in four-dimensional $U(1)$ there is an attractive interaction between flux tubes for $\beta$ approaching the phase transition.
\end{abstract}
 
\section{Introduction}

The formation of colour-electric flux tubes provides an intuitive physical picture for permanent confinement of quarks within hadrons. Moreover, this picture seems to be realized within QCD, as has widely been observed in lattice simulations. The mechanism leading to flux tube formation, however, is still subject to analytical and numerical investigations. A wide class of phenomenological models incorporates the existence of flux tubes: bag models \cite{bag}, models describing the QCD vacuum as a perfect dia-electric medium \cite{thirring}, and the physically appealing picture of a dual superconductor \cite{mandelstam}. The question arises which of these effective models are compatible with the numerical results of lattice simulations.

The shortcoming of such lattice calculations, however, is the fact that regarding sufficiently long flux tubes, i.~e.~large lattice sizes, becomes a numerically very difficult task. Thus many interesting questions could not be answered until now with sufficient precision. Numerical troubles become even more serious, if one considers sources in higher representations which is essential to discriminate among the various phenomenological models \cite{trottier}. Because of these difficulties it is convenient to consider gauge groups simpler than $SU(3)$, like $SU(2)$, or even compact QED ($U(1)$ gauge theory) which also exhibits confinement of charges and can be used as a prototype for investigating the formation of flux tubes.

In ref.~\cite{trottier} $SU(2)$ and $U(1)$ lattice gauge theory in three dimensions were used to calculate some features of flux tubes generated by sources in different representations. In particular, the cross section of flux tubes, the string tension and the spatial distribution of the energy density were investigated. The result was obtained that the cross section $\cal{A}$ is nearly independent of the representation $j$, and the string tension is proportional to the eigenvalue of the Casimir operator, i.~e.~$\sigma \propto j(j+1)$, which corresponds to the squared colour charge $Q_j^2$ (this has already been observed in previous lattice calculations). For multiple charges $Q=n\,e$ in three-dimensional $U(1)$ it was found that $\sigma \propto n^2$. The authors conjectured that their results should also be relevant for flux tubes in four-dimensional QCD; the similarity of results obtained in $SU(2)$ and $U(1)$ was interpreted as indirect support for the dual superconductor picture of confinement, which is well established for $U(1)$ lattice gauge theory.

However, as was also realized in ref.~\cite{trottier}, this situation is in conflict with some of the above mentioned phenomenological models. According to our opinion, even the dual superconductor scenario is not compatible with the $U(1)$ result $\sigma \propto n^2$ of ref.~\cite{trottier}. We expect that in a dual superconductor the string tension increases linearly with the number of flux quanta, apart from interaction between flux tubes: In a type-II superconductor a doubly charged flux tube will split into two separated flux tubes. The energy will be just twice the energy of a single flux tube, if their transverse distance from each other is sufficiently large. In a type-I superconductor there is a negative interaction energy for two flux tubes which should diminish the ratio of string tensions below two. 

In fact we will demonstrate that the string tension scales proportionally to the charge rather than the squared charge in the confinement phase of $U(1)$. In addition, investigating the interaction energy of doubly charged systems we will discuss the question if four-dimensional $U(1)$ behaves like a type-I or a type-II superconductor. The results concerning the flux tube cross-section and the energy obtained in ref.~\cite{trottier} on the other hand remind of the classical Coulomb case: The field strength is doubled everywhere in space, if we replace single by double charges, and the energy scales like the squared charge.

It is the main aim of this paper to test the behaviour of flux tubes with respect to the above mentioned attributes within the dual formulation of four-dimensional $U(1)$ lattice gauge theory. Simulating the dually transformed theory \cite{pr97} provides two great advantages compared to standard lattice simulations: The obtained signal does not decrease with the charge distance (the length of the flux tube), and a simulation with multiply charged sources can be performed with equal accuracy. Thanks to these features we will be able to perform a detailed investigation of single and double flux tubes in four-dimensional $U(1)$, and also analyze the reasons for the discrepancy of our results to those in ref.~\cite{trottier} for the three-dimensional case.

The remaining part of this article is organized as follows: In section 2 we shortly review the techniques used for the simulations. They are based on the method introduced in ref.~\cite{pr97}, in addition we explain how to implement periodically closed flux tubes and doubly charged sources, and how to calculate the free energy. Then, in section 3, we present our results for flux tubes between static (double) charges in four-dimensional $U(1)$. In section 4 we calculate the energy of periodically closed flux tubes (torelons). After a detailed analysis of the three-dimensional case in section 5, we finally turn to an investigation of the interaction between flux tubes in the four-dimensional theory in section 6.

\section{Methods}

We are interested in the expectation values of physical observables ${\cal O}$ in the presence of a static charge pair at distance $d$, which are determined by
\begin{equation} \label{correlation}
\langle {\cal O}(x) \rangle_{Q\bar{Q}} = \frac{\langle L(0) L^+(d) {\cal O}(x) \rangle}{\langle L(0) L^+(d) \rangle} - \langle {\cal O} \rangle,
\end{equation}
where $L$ denotes the Polyakov loop $L(\vec{r})=\prod_{k=1}^{N_{t}}U_{x=(\vec r,ka),\mu=4}$ and the angle brackets mean the averaging according to the path integral using the standard Wilson action
\begin{equation} \label{action}
S_W = \beta \sum_\mathit{plaquettes} [1-\cos d \theta].
\end{equation}
$\beta=1/e^2$ is the inverse coupling, and $d \theta$ is the discretized exterior derivative of the phases $\theta$ of the link variables $U_{x,\mu}=e^{i\theta_{x,\mu}}$ and is assigned to a plaquette of size $a^2$. For convenience we choose the notation of lattice differential forms.

Each of the three expectation values appearing on the right hand side of (\ref{correlation}) can be dually transformed as discussed in ref.~\cite{pr97}. The path integral including two external charges (${\cal L_+}$ and ${\cal L_-}$ denote the world-lines of a static charge pair) reads
\begin{equation}\label{wje}
Z_{Q\bar{Q}} = \prod \int^{\pi}_{-\pi} d \theta \, \exp\{- \beta \sum
(1 - \cos d\theta)\} \prod_{\cal L_+} e^{-i \theta} \prod_{\cal L_-} 
e^{i \theta} \, = \, (2 \pi)^{4N} \prod \sum_{^*l} \, \prod 
e^{-\beta} I_{\parallel d {}^*l + {}^*n \parallel}(\beta),
\end{equation}
where the first product on the right hand side is performed over all links of the dual lattice, and the Boltzmann factor is a product of modified Bessel functions over the dual plaquettes. $^*l$ is an integer valued 1-form. It is kept as dynamical variable and updated in simulations, while the 2-form $^*n$ is fixed to be equal to 1 on the minimal straight area between the two Polyakov loops, and zero everywhere else. If one interprets the dually transformed partition function as a certain limit of a dual Higgs model \cite{froehlich}, $n$ has the meaning of a dual Dirac sheet, and $^*k = d {}^*l + {}^*n$ is the field strength. For the $U(1)$ path integral without external sources one obtains the same as expr.~(\ref{wje}), but with $^*n \equiv 0$.

Further, we perform the duality transformation for the numerator in the correlation function (\ref{correlation}). For the determination of the flux $F_{\Box_0}$ through a given plaquette ${\Box_0}$, we have to insert the operator $F_{\Box_0} = \sqrt{\beta} \sin d \theta_{\Box_0}$ \cite{pr97, pl95} into the path integral in addition to the Polyakov loops. One obtains
\begin{eqnarray}
\left\langle F_{\Box_0}\phantom{^2}\hspace{-1.5mm} \right\rangle_{Q\bar{Q}} \;\; = \;\; Z^{-1}_{Q\bar{Q}} \;\; \left\langle \; \Im \; \prod_{\cal L_+} e^{-i \theta} \prod_{\cal L_-} e^{i \theta} \; \sqrt{\beta} \sin d \theta_{\Box_0} \; \right\rangle \;\; = \;\; \left\langle \; \frac{1}{\sqrt{\beta}} {}^*k_{\Box_0} \; \right\rangle_{Q\bar{Q}}^{dual},
\end{eqnarray}
where the angle brackets in the final expression denote averaging over the dual degrees of freedom according to partition function (\ref{wje}), and $^*k = d {}^*l + {}^*n$. The vacuum expectation value of the field strength $F$ is of course zero. For the square of the field strength, which we need for determining the local energy and action density, we take the usual operator $F^2_{\Box_0}/2 = \beta \cos d \theta_{\Box_0}$\footnote{In Minkowski space the electric energy density is then given by $E_i^2/2 = F^2_{\Box_{i4}}/2$, the magnetic energy density by $B_i^2/2 = -F^2_{\Box_{jk}}/2$.}. In the dual formulation this becomes
\begin{equation}\label{cos}
\frac{1}{2} \; \left\langle F^2_{\Box_0}\right\rangle_{Q\bar{Q}} \;\; = \;\; \left\langle \; \beta \frac{I^{\prime}_{{{}^*k}_{\Box_0}}(\beta)}{I_{{{}^*k}_{\Box_0}}(\beta)} \; \right\rangle_{Q\bar{Q}}^{dual} \;\; - \;\; \left\langle  \; \beta \frac{I^{\prime}_{{{}^*k}}(\beta)}{I_{{{}^*k}}(\beta)} \; \right\rangle_{vac}^{dual}.
\end{equation} 
An identity equivalent to (\ref{cos}) was already derived in ref.~\cite{greensite} for calculating the electromagnetic energy density for flux tubes in three-dimensional $U(1)$ lattice gauge theory, using the polymer formulation of the path integral.

As pointed out in ref.~\cite{pr97}, the correspondence of the dually transformed $U(1)$ path integral to a limit of a dual Higgs model can qualitatively explain the formation of a flux tube (as product of a fluctuating string), but it does not help very much for interpreting the $U(1)$ data in terms of a classical dual superconductor. Nevertheless, the above introduced dual formulation of expectation values in the presence of external charges provides a very efficient tool for flux tube simulations for several reasons.

Contrary to standard lattice simulations, the confinement phase is the weakly coupled one in the dual theory, therefore there are less quantum fluctuations. Even more important is the fact that it is not necessary to project the charge--anticharge state out of the vacuum: On a lattice of same size charge pairs with arbitrary distance can be simulated with equal accuracy. For each charge distance $d$ a separate simulation has to be performed, and also for the subtraction of the vacuum expectation value in (\ref{correlation}) resp.~(\ref{cos}) a distinct run is needed. The main advantage for the aims of this paper lies in the implementation of doubly charged flux tubes by just setting the integer field $n$ on the dual Dirac sheet to 2 instead of 1. 

Our typical measurements are taken from $5\cdot10^5 - 2\cdot10^6$ configurations (depending on the lattice size), where each 10th configuration is evaluated. The updating procedure uses a standard Metropolis algorithm; for determining the errorbars we took blocks of 100 evaluated configurations. We investigated the whole range of inverse coupling $\beta$ in the four-dimensional theory. To compare the results in the confinement phase to those in the Coulomb phase, we focus on simulations at $\beta=0.96$ and $\beta=1.10$. In three-dimensional lattice $U(1)$, where there is no phase transition to a deconfined phase, we also analyzed the $\beta$-dependence of the string tension.

For the required spatial size of the lattice there is the simple requirement that its spatial extent has to be greater than the extent of the considered flux tube both in longitudinal and in transverse direction in order to avoid finite size effects. For our simulations we used a lattice size of $8^3 \times 32$ for charge distances $d=1a - 5a$, $12^3 \times 32$ for simulating $d=5a - 8a$, and $16^3 \times 32$ for $d=8a - 12a$. Further we performed some simulations on larger lattices; the largest investigated charge distance was $d=22a$ (on a $26^3 \times 32$ lattice). Finite temperature effects may be neglected if the temporal extent of the lattice exceeds the length of the flux tube (note that enlarging the temporal extent does not lead to a fall-off in the signal) \cite{pr97}. A comparison between results from standard and from dual simulations has been performed in ref.~\cite{stl}.

It turns out to be very convenient for our purposes to study flux tubes not only between static charges but also wrapping around the periodic lattice. These closed flux tubes (``torelons'') are free of end effects and allow for a more reliable calculation of the string tension as well as the contribution from string fluctuations \cite{torelon}. In standard simulations such torelons can be implemented by spatial Polyakov loops which are sufficiently separated in time direction. In the dual formulation it is straightforward to ``remove the sources'': The sheet defined by $n=1$ (resp.~2 for a double flux tube) is winding around the lattice both in temporal and longitudinal direction, and $\delta n=0$ everywhere. In such a simulation the flux tube is not fixed any more and may move in transverse directions. Therefore we do not calculate field profiles, but just measure the energy of this state compared to the vacuum.

Finally we have to specify what we mean by ``energy''. One approach is to sum up the spatial energy density, calculated from the expectation values of the squared fields. This is a very difficult task in standard lattice simulations, but within the dual simulation we are able to determine the total field energy as a function of the flux tube length. On the other hand, the usual ``potential'' is the zero temperature limit of the free energy relative to the vacuum, calculated from the correlation function between two Polyakov loops. Because this corresponds to a ratio of dual partition functions, and not to a dual expectation value, it cannot be obtained directly by a dual simulation. Nevertheless, we may relate the derivative of the free energy with respect to $\beta$ to the average total action of the system:
\begin{equation}\label{dfdb}
\frac{\partial F_{Q\bar{Q}}}{\partial \beta} = - \sum_{\Box} \left\langle \cos d\theta_{\Box} \right\rangle_{Q\bar{Q}} \Big|_{\beta}\,.
\end{equation}
Therefore it is possible to determine the free energy by a numerical integration over the average action relative to the vacuum as a function of $\beta$, starting from the well-known strong coupling behaviour $F_{Q(0)\bar{Q}(d)} \to - d \, \ln (\beta/2)$ as $\beta \to 0$. In the three-dimensional theory this procedure yields reliable results for the whole $\beta$-range, whereas in four-dimensional $U(1)$ integrating over the peak in the action density at the phase transition point ($\beta \approx 1.01$) leads to large errorbars in the Coulomb phase. Above the phase transition it is therefore preferable to perform the integration starting from the weak coupling limit of $U(1)$ gauge theory. It is obvious that the free energy for periodically closed flux tubes can be calculated in the same way.

We may recapitulate that by dual simulations we are able to investigate large charge distances (flux tube lengths) at very low temperatures. Simulating doubly charged systems is possible without any loss of accuracy, and by studying periodically closed flux tubes one avoids being disturbed by end effects. For our analysis we will consider both the total field energy and the free energy of flux tubes in $U(1)$ lattice gauge theory.

\section{Flux tubes between static double charges}

Usually a Wilson loop or a pair of Polyakov loops is used to generate a flux tube. In both cases the temporal extent has to be large enough to prevent strong contaminations by excited states respective finite temperature effects. What makes life difficult is the exponential fall-off of the signal with the temporal extent of loops. As already pointed out in the last section, in the dual formulation in principle there is no problem to go to arbitrary low temperatures. We represent static charges by Polyakov loops and calculate the time averages of the interesting observables. The quality of the signal is constant for arbitrary time extents and for multiply charged systems, and also for increasing charge distances (as long as we use a lattice of same size). In this section we will investigate the flux tube profile and the total field energy for a system of (double) charges at a distance $d$.

The connection between the flux tube cross section $\cal A$ and the string tension $\sigma$ was investigated in ref.~\cite{trottier}. Assuming that flux tube models of confinement are consistent with lattice simulations, the following relation should be valid for $U(1)$ gauge theory with charges $Q$ that are integer multiples of the coupling $e$:
\begin{equation}\label{ft}
\sigma_Q = \frac{Q^2}{2 {\cal A}_Q}.
\end{equation}
This means that if the cross section ${\cal A}$ is independent of the representation, the string tension will scale proportionally to the squared charge. Such a behaviour has been reported in ref.~\cite{trottier} for three-dimensional lattice gauge theories. In a large class of phenomenological models, however, $\cal{A}$ is expected to increase with the representation. We are going to investigate the dependence of the flux tube cross section on the charge $Q$ of the sources. For our analysis we will use the definition of $\cal{A}$ suggested in ref.~\cite{trottier}
\begin{equation}\label{adef}
{\cal A} = \frac{(\int \vec{E} d\vec{A})^2}{\int \vec{E}^2 dA} = \frac{Q^2}{\int \vec{E}^2 dA},
\end{equation}
where in the second expression we have exploited the exact validity of Gauss' law \cite{pr97, pl95}. Let us first compare the flux tube profiles for single and double charges. In fig.~\ref{prof} the longitudinal component $E_\|$ of the electric field as well as its square $E_\|^2$ are shown in the symmetry plane between charges as a function of transverse distance $R$ from the axis. Note that the distance between the charges is 22 lattice spacings. In the upper plot it can be seen that the field strength of the double flux tube on the axis is significantly lower than twice the field strength of a single flux tube: The electric flux seems to spread over a wider transverse area. The plot for the distribution of $E_\|^2$ looks very similar. We therefore expect that, calculating the cross sections according to definition (\ref{adef}), the double flux tube cross section will be significantly greater than that of a single flux tube, at least for this large charge distance. We show in fig.~\ref{area} that this is indeed the case. The ratio ${\cal A}_{double}/{\cal A}$ is plotted as a function of the distance between charges. For larger distances it is equal to 2 within the errorbars; the deviation at small distances seems to result from Coulombic contributions important for these distances.
\begin{figure}
\centerline{
\setlength{\unitlength}{0.1bp}
\special{!
/gnudict 40 dict def
gnudict begin
/Color false def
/Solid false def
/gnulinewidth 5.000 def
/vshift -33 def
/dl {10 mul} def
/hpt 31.5 def
/vpt 31.5 def
/M {moveto} bind def
/L {lineto} bind def
/R {rmoveto} bind def
/V {rlineto} bind def
/vpt2 vpt 2 mul def
/hpt2 hpt 2 mul def
/Lshow { currentpoint stroke M
  0 vshift R show } def
/Rshow { currentpoint stroke M
  dup stringwidth pop neg vshift R show } def
/Cshow { currentpoint stroke M
  dup stringwidth pop -2 div vshift R show } def
/DL { Color {setrgbcolor Solid {pop []} if 0 setdash }
 {pop pop pop Solid {pop []} if 0 setdash} ifelse } def
/BL { stroke gnulinewidth 2 mul setlinewidth } def
/AL { stroke gnulinewidth 2 div setlinewidth } def
/PL { stroke gnulinewidth setlinewidth } def
/LTb { BL [] 0 0 0 DL } def
/LTa { AL [1 dl 2 dl] 0 setdash 0 0 0 setrgbcolor } def
/LT0 { PL [] 0 1 0 DL } def
/LT1 { PL [4 dl 2 dl] 0 0 1 DL } def
/LT2 { PL [2 dl 3 dl] 1 0 0 DL } def
/LT3 { PL [1 dl 1.5 dl] 1 0 1 DL } def
/LT4 { PL [5 dl 2 dl 1 dl 2 dl] 0 1 1 DL } def
/LT5 { PL [4 dl 3 dl 1 dl 3 dl] 1 1 0 DL } def
/LT6 { PL [2 dl 2 dl 2 dl 4 dl] 0 0 0 DL } def
/LT7 { PL [2 dl 2 dl 2 dl 2 dl 2 dl 4 dl] 1 0.3 0 DL } def
/LT8 { PL [2 dl 2 dl 2 dl 2 dl 2 dl 2 dl 2 dl 4 dl] 0.5 0.5 0.5 DL } def
/P { stroke [] 0 setdash
  currentlinewidth 2 div sub M
  0 currentlinewidth V stroke } def
/D { stroke [] 0 setdash 2 copy vpt add M
  hpt neg vpt neg V hpt vpt neg V
  hpt vpt V hpt neg vpt V closepath stroke
  P } def
/A { stroke [] 0 setdash vpt sub M 0 vpt2 V
  currentpoint stroke M
  hpt neg vpt neg R hpt2 0 V stroke
  } def
/B { stroke [] 0 setdash 2 copy exch hpt sub exch vpt add M
  0 vpt2 neg V hpt2 0 V 0 vpt2 V
  hpt2 neg 0 V closepath stroke
  P } def
/C { stroke [] 0 setdash exch hpt sub exch vpt add M
  hpt2 vpt2 neg V currentpoint stroke M
  hpt2 neg 0 R hpt2 vpt2 V stroke } def
/T { stroke [] 0 setdash 2 copy vpt 1.12 mul add M
  hpt neg vpt -1.62 mul V
  hpt 2 mul 0 V
  hpt neg vpt 1.62 mul V closepath stroke
  P  } def
/S { 2 copy A C} def
end
}
\begin{picture}(3600,2160)(0,0)
\special{"
gnudict begin
gsave
50 50 translate
0.100 0.100 scale
0 setgray
/Helvetica findfont 100 scalefont setfont
newpath
-500.000000 -500.000000 translate
LTa
480 361 M
2937 0 V
538 251 M
0 1758 V
LTb
480 361 M
63 0 V
2874 0 R
-63 0 V
480 910 M
63 0 V
2874 0 R
-63 0 V
480 1460 M
63 0 V
2874 0 R
-63 0 V
480 2009 M
63 0 V
2874 0 R
-63 0 V
538 251 M
0 63 V
0 1695 R
0 -63 V
1113 251 M
0 63 V
0 1695 R
0 -63 V
1689 251 M
0 63 V
0 1695 R
0 -63 V
2265 251 M
0 63 V
0 1695 R
0 -63 V
2841 251 M
0 63 V
0 1695 R
0 -63 V
3417 251 M
0 63 V
0 1695 R
0 -63 V
480 251 M
2937 0 V
0 1758 V
-2937 0 V
480 251 L
LT0
3114 1846 M
180 0 V
538 1472 M
575 -305 V
1352 940 L
1689 692 L
1825 591 L
2166 454 L
99 -23 V
94 -24 V
255 -25 V
227 -16 V
71 1 V
69 -4 V
132 1 V
304 -6 V
0 6 V
LT1
3114 1746 M
180 0 V
538 1813 M
575 -216 V
239 -206 V
337 -337 V
1825 906 L
2166 634 L
99 -53 V
94 -50 V
255 -89 V
227 -34 V
71 -22 V
69 8 V
132 -17 V
304 -15 V
0 3 V
LT0
538 1472 P
1113 1167 P
1352 940 P
1689 692 P
1825 591 P
2166 454 P
2265 431 P
2359 407 P
2614 382 P
2841 366 P
2912 367 P
2981 363 P
3113 364 P
3417 358 P
3417 364 P
538 1439 M
0 66 V
-31 -66 R
62 0 V
-62 66 R
62 0 V
544 -357 R
0 38 V
-31 -38 R
62 0 V
-62 38 R
62 0 V
1352 925 M
0 31 V
-31 -31 R
62 0 V
-62 31 R
62 0 V
1689 680 M
0 24 V
-31 -24 R
62 0 V
-62 24 R
62 0 V
1825 584 M
0 13 V
-31 -13 R
62 0 V
-62 13 R
62 0 V
2166 445 M
0 18 V
-31 -18 R
62 0 V
-62 18 R
62 0 V
68 -41 R
0 17 V
-31 -17 R
62 0 V
-62 17 R
62 0 V
63 -38 R
0 11 V
-31 -11 R
62 0 V
-62 11 R
62 0 V
224 -35 R
0 10 V
-31 -10 R
62 0 V
-62 10 R
62 0 V
196 -26 R
0 11 V
-31 -11 R
62 0 V
-62 11 R
62 0 V
40 -9 R
0 7 V
-31 -7 R
62 0 V
-62 7 R
62 0 V
38 -13 R
0 11 V
-31 -11 R
62 0 V
-62 11 R
62 0 V
101 -7 R
0 6 V
-31 -6 R
62 0 V
-62 6 R
62 0 V
273 -13 R
0 8 V
-31 -8 R
62 0 V
-62 8 R
62 0 V
-31 -1 R
0 7 V
-31 -7 R
62 0 V
-62 7 R
62 0 V
LT0
538 1813 P
1113 1597 P
1352 1391 P
1689 1054 P
1825 906 P
2166 634 P
2265 581 P
2359 531 P
2614 442 P
2841 408 P
2912 386 P
2981 394 P
3113 377 P
3417 362 P
3417 365 P
538 1780 M
0 66 V
-31 -66 R
62 0 V
-62 66 R
62 0 V
544 -268 R
0 38 V
-31 -38 R
62 0 V
-62 38 R
62 0 V
208 -241 R
0 31 V
-31 -31 R
62 0 V
-62 31 R
62 0 V
306 -364 R
0 24 V
-31 -24 R
62 0 V
-62 24 R
62 0 V
1825 899 M
0 13 V
-31 -13 R
62 0 V
-62 13 R
62 0 V
2166 624 M
0 19 V
-31 -19 R
62 0 V
-62 19 R
62 0 V
68 -70 R
0 16 V
-31 -16 R
62 0 V
-62 16 R
62 0 V
63 -63 R
0 11 V
-31 -11 R
62 0 V
-62 11 R
62 0 V
224 -99 R
0 9 V
-31 -9 R
62 0 V
-62 9 R
62 0 V
196 -45 R
0 11 V
-31 -11 R
62 0 V
-62 11 R
62 0 V
40 -31 R
0 8 V
-31 -8 R
62 0 V
-62 8 R
62 0 V
38 -2 R
0 11 V
-31 -11 R
62 0 V
-62 11 R
62 0 V
101 -25 R
0 6 V
-31 -6 R
62 0 V
-62 6 R
62 0 V
273 -22 R
0 8 V
-31 -8 R
62 0 V
-62 8 R
62 0 V
-31 -4 R
0 6 V
-31 -6 R
62 0 V
-62 6 R
62 0 V
stroke
grestore
end
showpage
}
\put(3054,1746){\makebox(0,0)[r]{$E_{\| double}$}}
\put(3054,1846){\makebox(0,0)[r]{$E_\|$}}
\put(1948,2109){\makebox(0,0){$d=22a$}}
\put(1948,51){\makebox(0,0){R [a]}}
\put(3417,151){\makebox(0,0){5}}
\put(2841,151){\makebox(0,0){4}}
\put(2265,151){\makebox(0,0){3}}
\put(1689,151){\makebox(0,0){2}}
\put(1113,151){\makebox(0,0){1}}
\put(538,151){\makebox(0,0){0}}
\put(420,2009){\makebox(0,0)[r]{0.15}}
\put(420,1460){\makebox(0,0)[r]{0.1}}
\put(420,910){\makebox(0,0)[r]{0.05}}
\put(420,361){\makebox(0,0)[r]{0}}
\end{picture}}
\vspace{1cm}
\centerline{
\setlength{\unitlength}{0.1bp}
\special{!
/gnudict 40 dict def
gnudict begin
/Color false def
/Solid false def
/gnulinewidth 5.000 def
/vshift -33 def
/dl {10 mul} def
/hpt 31.5 def
/vpt 31.5 def
/M {moveto} bind def
/L {lineto} bind def
/R {rmoveto} bind def
/V {rlineto} bind def
/vpt2 vpt 2 mul def
/hpt2 hpt 2 mul def
/Lshow { currentpoint stroke M
  0 vshift R show } def
/Rshow { currentpoint stroke M
  dup stringwidth pop neg vshift R show } def
/Cshow { currentpoint stroke M
  dup stringwidth pop -2 div vshift R show } def
/DL { Color {setrgbcolor Solid {pop []} if 0 setdash }
 {pop pop pop Solid {pop []} if 0 setdash} ifelse } def
/BL { stroke gnulinewidth 2 mul setlinewidth } def
/AL { stroke gnulinewidth 2 div setlinewidth } def
/PL { stroke gnulinewidth setlinewidth } def
/LTb { BL [] 0 0 0 DL } def
/LTa { AL [1 dl 2 dl] 0 setdash 0 0 0 setrgbcolor } def
/LT0 { PL [] 0 1 0 DL } def
/LT1 { PL [4 dl 2 dl] 0 0 1 DL } def
/LT2 { PL [2 dl 3 dl] 1 0 0 DL } def
/LT3 { PL [1 dl 1.5 dl] 1 0 1 DL } def
/LT4 { PL [5 dl 2 dl 1 dl 2 dl] 0 1 1 DL } def
/LT5 { PL [4 dl 3 dl 1 dl 3 dl] 1 1 0 DL } def
/LT6 { PL [2 dl 2 dl 2 dl 4 dl] 0 0 0 DL } def
/LT7 { PL [2 dl 2 dl 2 dl 2 dl 2 dl 4 dl] 1 0.3 0 DL } def
/LT8 { PL [2 dl 2 dl 2 dl 2 dl 2 dl 2 dl 2 dl 4 dl] 0.5 0.5 0.5 DL } def
/P { stroke [] 0 setdash
  currentlinewidth 2 div sub M
  0 currentlinewidth V stroke } def
/D { stroke [] 0 setdash 2 copy vpt add M
  hpt neg vpt neg V hpt vpt neg V
  hpt vpt V hpt neg vpt V closepath stroke
  P } def
/A { stroke [] 0 setdash vpt sub M 0 vpt2 V
  currentpoint stroke M
  hpt neg vpt neg R hpt2 0 V stroke
  } def
/B { stroke [] 0 setdash 2 copy exch hpt sub exch vpt add M
  0 vpt2 neg V hpt2 0 V 0 vpt2 V
  hpt2 neg 0 V closepath stroke
  P } def
/C { stroke [] 0 setdash exch hpt sub exch vpt add M
  hpt2 vpt2 neg V currentpoint stroke M
  hpt2 neg 0 R hpt2 vpt2 V stroke } def
/T { stroke [] 0 setdash 2 copy vpt 1.12 mul add M
  hpt neg vpt -1.62 mul V
  hpt 2 mul 0 V
  hpt neg vpt 1.62 mul V closepath stroke
  P  } def
/S { 2 copy A C} def
end
}
\begin{picture}(3600,2160)(0,0)
\special{"
gnudict begin
gsave
50 50 translate
0.100 0.100 scale
0 setgray
/Helvetica findfont 100 scalefont setfont
newpath
-500.000000 -500.000000 translate
LTa
480 319 M
2937 0 V
538 251 M
0 1758 V
LTb
480 319 M
63 0 V
2874 0 R
-63 0 V
480 657 M
63 0 V
2874 0 R
-63 0 V
480 995 M
63 0 V
2874 0 R
-63 0 V
480 1333 M
63 0 V
2874 0 R
-63 0 V
480 1671 M
63 0 V
2874 0 R
-63 0 V
480 2009 M
63 0 V
2874 0 R
-63 0 V
538 251 M
0 63 V
0 1695 R
0 -63 V
1113 251 M
0 63 V
0 1695 R
0 -63 V
1689 251 M
0 63 V
0 1695 R
0 -63 V
2265 251 M
0 63 V
0 1695 R
0 -63 V
2841 251 M
0 63 V
0 1695 R
0 -63 V
3417 251 M
0 63 V
0 1695 R
0 -63 V
480 251 M
2937 0 V
0 1758 V
-2937 0 V
480 251 L
LT0
3114 1846 M
180 0 V
538 1400 M
575 -279 V
1352 913 L
1689 655 L
136 -91 V
2166 426 L
99 -26 V
94 -16 V
255 -38 V
367 -17 V
-140 4 V
71 -3 V
201 -4 V
304 -1 V
0 1 V
LT1
3114 1746 M
180 0 V
538 1787 M
575 -204 V
239 -200 V
337 -341 V
1825 895 L
2166 612 L
99 -45 V
94 -50 V
2614 416 L
367 -68 V
-140 17 V
71 -5 V
201 -16 V
304 -23 V
0 3 V
LT0
538 1400 P
1113 1121 P
1352 913 P
1689 655 P
1825 564 P
2166 426 P
2265 400 P
2359 384 P
2614 346 P
2981 329 P
2841 333 P
2912 330 P
3113 326 P
3417 325 P
3417 326 P
538 1372 M
0 57 V
-31 -57 R
62 0 V
-62 57 R
62 0 V
544 -327 R
0 38 V
-31 -38 R
62 0 V
-62 38 R
62 0 V
1352 898 M
0 30 V
-31 -30 R
62 0 V
-62 30 R
62 0 V
1689 645 M
0 21 V
-31 -21 R
62 0 V
-62 21 R
62 0 V
1825 558 M
0 13 V
-31 -13 R
62 0 V
-62 13 R
62 0 V
2166 417 M
0 17 V
-31 -17 R
62 0 V
-62 17 R
62 0 V
68 -43 R
0 17 V
-31 -17 R
62 0 V
-62 17 R
62 0 V
63 -31 R
0 13 V
-31 -13 R
62 0 V
-62 13 R
62 0 V
224 -49 R
0 10 V
-31 -10 R
62 0 V
-62 10 R
62 0 V
336 -27 R
0 11 V
-31 -11 R
62 0 V
-62 11 R
62 0 V
-171 -7 R
0 11 V
-31 -11 R
62 0 V
-62 11 R
62 0 V
40 -13 R
0 8 V
-31 -8 R
62 0 V
-62 8 R
62 0 V
170 -12 R
0 8 V
-31 -8 R
62 0 V
-62 8 R
62 0 V
273 -9 R
0 7 V
-31 -7 R
62 0 V
-62 7 R
62 0 V
-31 -7 R
0 9 V
-31 -9 R
62 0 V
-62 9 R
62 0 V
LT0
538 1787 P
1113 1583 P
1352 1383 P
1689 1042 P
1825 895 P
2166 612 P
2265 567 P
2359 517 P
2614 416 P
2981 348 P
2841 365 P
2912 360 P
3113 344 P
3417 321 P
3417 324 P
538 1759 M
0 56 V
-31 -56 R
62 0 V
-62 56 R
62 0 V
544 -251 R
0 37 V
-31 -37 R
62 0 V
-62 37 R
62 0 V
208 -232 R
0 29 V
-31 -29 R
62 0 V
-62 29 R
62 0 V
306 -366 R
0 21 V
-31 -21 R
62 0 V
-62 21 R
62 0 V
1825 889 M
0 13 V
-31 -13 R
62 0 V
-62 13 R
62 0 V
2166 603 M
0 17 V
-31 -17 R
62 0 V
-62 17 R
62 0 V
68 -61 R
0 17 V
-31 -17 R
62 0 V
-62 17 R
62 0 V
63 -66 R
0 13 V
-31 -13 R
62 0 V
-62 13 R
62 0 V
2614 410 M
0 11 V
-31 -11 R
62 0 V
-62 11 R
62 0 V
336 -78 R
0 10 V
-31 -10 R
62 0 V
-62 10 R
62 0 V
-171 7 R
0 11 V
-31 -11 R
62 0 V
-62 11 R
62 0 V
40 -15 R
0 8 V
-31 -8 R
62 0 V
-62 8 R
62 0 V
170 -24 R
0 8 V
-31 -8 R
62 0 V
-62 8 R
62 0 V
273 -30 R
0 7 V
-31 -7 R
62 0 V
-62 7 R
62 0 V
-31 -5 R
0 9 V
-31 -9 R
62 0 V
-62 9 R
62 0 V
stroke
grestore
end
showpage
}
\put(3054,1746){\makebox(0,0)[r]{$E^2_{\| double}$}}
\put(3054,1846){\makebox(0,0)[r]{$E^2_\|$}}
\put(1948,2109){\makebox(0,0){$d=22a$}}
\put(1948,51){\makebox(0,0){R [a]}}
\put(3417,151){\makebox(0,0){5}}
\put(2841,151){\makebox(0,0){4}}
\put(2265,151){\makebox(0,0){3}}
\put(1689,151){\makebox(0,0){2}}
\put(1113,151){\makebox(0,0){1}}
\put(538,151){\makebox(0,0){0}}
\put(420,2009){\makebox(0,0)[r]{0.25}}
\put(420,1671){\makebox(0,0)[r]{0.2}}
\put(420,1333){\makebox(0,0)[r]{0.15}}
\put(420,995){\makebox(0,0)[r]{0.1}}
\put(420,657){\makebox(0,0)[r]{0.05}}
\put(420,319){\makebox(0,0)[r]{0}}
\end{picture}}
\caption{\label{prof}Flux tube profile of longitudinal electric field ($E_\|$) and its square ($E_\|^2$) in the symmetry plane between single (solid line) and double charges (dashed line) at a distance $d=22a$ for $\beta=0.96$.}
\end{figure}
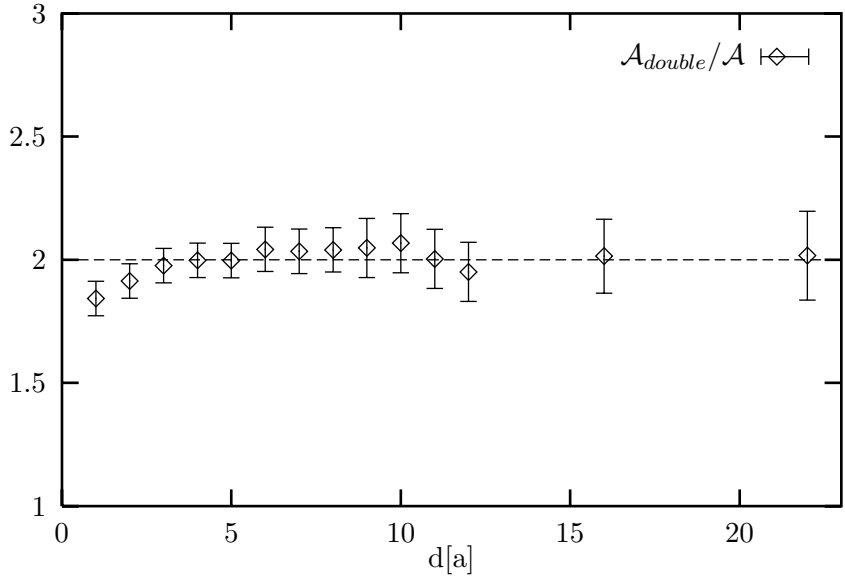
\begin{figure}
\centerline{
\setlength{\unitlength}{0.1bp}
\special{!
/gnudict 40 dict def
gnudict begin
/Color false def
/Solid false def
/gnulinewidth 5.000 def
/vshift -33 def
/dl {10 mul} def
/hpt 31.5 def
/vpt 31.5 def
/M {moveto} bind def
/L {lineto} bind def
/R {rmoveto} bind def
/V {rlineto} bind def
/vpt2 vpt 2 mul def
/hpt2 hpt 2 mul def
/Lshow { currentpoint stroke M
  0 vshift R show } def
/Rshow { currentpoint stroke M
  dup stringwidth pop neg vshift R show } def
/Cshow { currentpoint stroke M
  dup stringwidth pop -2 div vshift R show } def
/DL { Color {setrgbcolor Solid {pop []} if 0 setdash }
 {pop pop pop Solid {pop []} if 0 setdash} ifelse } def
/BL { stroke gnulinewidth 2 mul setlinewidth } def
/AL { stroke gnulinewidth 2 div setlinewidth } def
/PL { stroke gnulinewidth setlinewidth } def
/LTb { BL [] 0 0 0 DL } def
/LTa { AL [1 dl 2 dl] 0 setdash 0 0 0 setrgbcolor } def
/LT0 { PL [] 0 1 0 DL } def
/LT1 { PL [4 dl 2 dl] 0 0 1 DL } def
/LT2 { PL [2 dl 3 dl] 1 0 0 DL } def
/LT3 { PL [1 dl 1.5 dl] 1 0 1 DL } def
/LT4 { PL [5 dl 2 dl 1 dl 2 dl] 0 1 1 DL } def
/LT5 { PL [4 dl 3 dl 1 dl 3 dl] 1 1 0 DL } def
/LT6 { PL [2 dl 2 dl 2 dl 4 dl] 0 0 0 DL } def
/LT7 { PL [2 dl 2 dl 2 dl 2 dl 2 dl 4 dl] 1 0.3 0 DL } def
/LT8 { PL [2 dl 2 dl 2 dl 2 dl 2 dl 2 dl 2 dl 4 dl] 0.5 0.5 0.5 DL } def
/P { stroke [] 0 setdash
  currentlinewidth 2 div sub M
  0 currentlinewidth V stroke } def
/D { stroke [] 0 setdash 2 copy vpt add M
  hpt neg vpt neg V hpt vpt neg V
  hpt vpt V hpt neg vpt V closepath stroke
  P } def
/A { stroke [] 0 setdash vpt sub M 0 vpt2 V
  currentpoint stroke M
  hpt neg vpt neg R hpt2 0 V stroke
  } def
/B { stroke [] 0 setdash 2 copy exch hpt sub exch vpt add M
  0 vpt2 neg V hpt2 0 V 0 vpt2 V
  hpt2 neg 0 V closepath stroke
  P } def
/C { stroke [] 0 setdash exch hpt sub exch vpt add M
  hpt2 vpt2 neg V currentpoint stroke M
  hpt2 neg 0 R hpt2 vpt2 V stroke } def
/T { stroke [] 0 setdash 2 copy vpt 1.12 mul add M
  hpt neg vpt -1.62 mul V
  hpt 2 mul 0 V
  hpt neg vpt 1.62 mul V closepath stroke
  P  } def
/S { 2 copy A C} def
end
}
\begin{picture}(3600,2160)(0,0)
\special{"
gnudict begin
gsave
50 50 translate
0.100 0.100 scale
0 setgray
/Helvetica findfont 100 scalefont setfont
newpath
-500.000000 -500.000000 translate
LTa
480 251 M
0 1858 V
LTb
480 251 M
63 0 V
2874 0 R
-63 0 V
480 716 M
63 0 V
2874 0 R
-63 0 V
480 1180 M
63 0 V
2874 0 R
-63 0 V
480 1645 M
63 0 V
2874 0 R
-63 0 V
480 2109 M
63 0 V
2874 0 R
-63 0 V
480 251 M
0 63 V
0 1795 R
0 -63 V
1118 251 M
0 63 V
0 1795 R
0 -63 V
1757 251 M
0 63 V
0 1795 R
0 -63 V
2395 251 M
0 63 V
0 1795 R
0 -63 V
3034 251 M
0 63 V
0 1795 R
0 -63 V
480 251 M
2937 0 V
0 1858 V
-2937 0 V
480 251 L
LT0
3174 1946 D
608 1034 D
735 1100 D
863 1158 D
991 1178 D
1118 1177 D
1246 1219 D
1374 1212 D
1502 1217 D
1629 1225 D
1757 1243 D
1885 1183 D
2012 1134 D
2523 1194 D
3289 1196 D
3114 1946 M
180 0 V
-180 31 R
0 -62 V
180 62 R
0 -62 V
608 969 M
0 130 V
577 969 M
62 0 V
-62 130 R
62 0 V
96 -64 R
0 130 V
704 1035 M
62 0 V
-62 130 R
62 0 V
97 -72 R
0 130 V
832 1093 M
62 0 V
-62 130 R
62 0 V
97 -110 R
0 130 V
960 1113 M
62 0 V
-62 130 R
62 0 V
96 -131 R
0 130 V
-31 -130 R
62 0 V
-62 130 R
62 0 V
97 -106 R
0 167 V
-31 -167 R
62 0 V
-62 167 R
62 0 V
97 -175 R
0 168 V
-31 -168 R
62 0 V
-62 168 R
62 0 V
97 -162 R
0 167 V
-31 -167 R
62 0 V
-62 167 R
62 0 V
96 -188 R
0 223 V
-31 -223 R
62 0 V
-62 223 R
62 0 V
97 -205 R
0 223 V
-31 -223 R
62 0 V
-62 223 R
62 0 V
97 -282 R
0 223 V
-31 -223 R
62 0 V
-62 223 R
62 0 V
96 -272 R
0 223 V
-31 -223 R
62 0 V
-62 223 R
62 0 V
480 -192 R
0 279 V
-31 -279 R
62 0 V
-62 279 R
62 0 V
735 -305 R
0 335 V
-31 -335 R
62 0 V
-62 335 R
62 0 V
LT1
480 1180 M
30 0 V
29 0 V
30 0 V
30 0 V
29 0 V
30 0 V
30 0 V
29 0 V
30 0 V
30 0 V
29 0 V
30 0 V
30 0 V
29 0 V
30 0 V
30 0 V
29 0 V
30 0 V
30 0 V
29 0 V
30 0 V
30 0 V
29 0 V
30 0 V
30 0 V
29 0 V
30 0 V
30 0 V
29 0 V
30 0 V
30 0 V
29 0 V
30 0 V
30 0 V
29 0 V
30 0 V
30 0 V
29 0 V
30 0 V
30 0 V
29 0 V
30 0 V
30 0 V
29 0 V
30 0 V
30 0 V
29 0 V
30 0 V
30 0 V
29 0 V
30 0 V
30 0 V
29 0 V
30 0 V
30 0 V
29 0 V
30 0 V
30 0 V
29 0 V
30 0 V
30 0 V
29 0 V
30 0 V
30 0 V
29 0 V
30 0 V
30 0 V
29 0 V
30 0 V
30 0 V
29 0 V
30 0 V
30 0 V
29 0 V
30 0 V
30 0 V
29 0 V
30 0 V
30 0 V
29 0 V
30 0 V
30 0 V
29 0 V
30 0 V
30 0 V
29 0 V
30 0 V
30 0 V
29 0 V
30 0 V
30 0 V
29 0 V
30 0 V
30 0 V
29 0 V
30 0 V
30 0 V
29 0 V
30 0 V
stroke
grestore
end
showpage
}
\put(3054,1946){\makebox(0,0)[r]{${\cal A}_{double}/{\cal A}$}}
\put(1948,51){\makebox(0,0){d[a]}}
\put(3034,151){\makebox(0,0){20}}
\put(2395,151){\makebox(0,0){15}}
\put(1757,151){\makebox(0,0){10}}
\put(1118,151){\makebox(0,0){5}}
\put(480,151){\makebox(0,0){0}}
\put(420,2109){\makebox(0,0)[r]{3}}
\put(420,1645){\makebox(0,0)[r]{2.5}}
\put(420,1180){\makebox(0,0)[r]{2}}
\put(420,716){\makebox(0,0)[r]{1.5}}
\put(420,251){\makebox(0,0)[r]{1}}
\end{picture}}
\caption{\label{area}Ratio of cross sections of a double and a single flux tube in the symmetry plane as a function of charge distance $d$ for $\beta=0.96$.}
\end{figure}

In terms of the flux tube model (\ref{ft}) this increase of the cross section indicates that the string tension increases less rapidly than the squared charge. According to the ratio shown in fig.~\ref{area} we should expect $\sigma \propto Q$. We postpone a further discussion of this question to the next section, because the string tension should be extracted from the potential, which differs from the electromagnetic field energy.

\begin{figure}
\centerline{$\beta=0.96$}
\centerline{
\setlength{\unitlength}{0.1bp}
\special{!
/gnudict 40 dict def
gnudict begin
/Color false def
/Solid false def
/gnulinewidth 5.000 def
/vshift -33 def
/dl {10 mul} def
/hpt 31.5 def
/vpt 31.5 def
/M {moveto} bind def
/L {lineto} bind def
/R {rmoveto} bind def
/V {rlineto} bind def
/vpt2 vpt 2 mul def
/hpt2 hpt 2 mul def
/Lshow { currentpoint stroke M
  0 vshift R show } def
/Rshow { currentpoint stroke M
  dup stringwidth pop neg vshift R show } def
/Cshow { currentpoint stroke M
  dup stringwidth pop -2 div vshift R show } def
/DL { Color {setrgbcolor Solid {pop []} if 0 setdash }
 {pop pop pop Solid {pop []} if 0 setdash} ifelse } def
/BL { stroke gnulinewidth 2 mul setlinewidth } def
/AL { stroke gnulinewidth 2 div setlinewidth } def
/PL { stroke gnulinewidth setlinewidth } def
/LTb { BL [] 0 0 0 DL } def
/LTa { AL [1 dl 2 dl] 0 setdash 0 0 0 setrgbcolor } def
/LT0 { PL [] 0 1 0 DL } def
/LT1 { PL [4 dl 2 dl] 0 0 1 DL } def
/LT2 { PL [2 dl 3 dl] 1 0 0 DL } def
/LT3 { PL [1 dl 1.5 dl] 1 0 1 DL } def
/LT4 { PL [5 dl 2 dl 1 dl 2 dl] 0 1 1 DL } def
/LT5 { PL [4 dl 3 dl 1 dl 3 dl] 1 1 0 DL } def
/LT6 { PL [2 dl 2 dl 2 dl 4 dl] 0 0 0 DL } def
/LT7 { PL [2 dl 2 dl 2 dl 2 dl 2 dl 4 dl] 1 0.3 0 DL } def
/LT8 { PL [2 dl 2 dl 2 dl 2 dl 2 dl 2 dl 2 dl 4 dl] 0.5 0.5 0.5 DL } def
/P { stroke [] 0 setdash
  currentlinewidth 2 div sub M
  0 currentlinewidth V stroke } def
/D { stroke [] 0 setdash 2 copy vpt add M
  hpt neg vpt neg V hpt vpt neg V
  hpt vpt V hpt neg vpt V closepath stroke
  P } def
/A { stroke [] 0 setdash vpt sub M 0 vpt2 V
  currentpoint stroke M
  hpt neg vpt neg R hpt2 0 V stroke
  } def
/B { stroke [] 0 setdash 2 copy exch hpt sub exch vpt add M
  0 vpt2 neg V hpt2 0 V 0 vpt2 V
  hpt2 neg 0 V closepath stroke
  P } def
/C { stroke [] 0 setdash exch hpt sub exch vpt add M
  hpt2 vpt2 neg V currentpoint stroke M
  hpt2 neg 0 R hpt2 vpt2 V stroke } def
/T { stroke [] 0 setdash 2 copy vpt 1.12 mul add M
  hpt neg vpt -1.62 mul V
  hpt 2 mul 0 V
  hpt neg vpt 1.62 mul V closepath stroke
  P  } def
/S { 2 copy A C} def
end
}
\begin{picture}(3600,2160)(0,0)
\special{"
gnudict begin
gsave
50 50 translate
0.100 0.100 scale
0 setgray
/Helvetica findfont 100 scalefont setfont
newpath
-500.000000 -500.000000 translate
LTa
480 251 M
2937 0 V
480 251 M
0 1858 V
LTb
480 251 M
63 0 V
2874 0 R
-63 0 V
480 716 M
63 0 V
2874 0 R
-63 0 V
480 1180 M
63 0 V
2874 0 R
-63 0 V
480 1645 M
63 0 V
2874 0 R
-63 0 V
480 2109 M
63 0 V
2874 0 R
-63 0 V
480 251 M
0 63 V
0 1795 R
0 -63 V
1118 251 M
0 63 V
0 1795 R
0 -63 V
1757 251 M
0 63 V
0 1795 R
0 -63 V
2395 251 M
0 63 V
0 1795 R
0 -63 V
3034 251 M
0 63 V
0 1795 R
0 -63 V
480 251 M
2937 0 V
0 1858 V
-2937 0 V
480 251 L
LT0
480 251 P
608 311 P
735 361 P
863 394 P
991 427 P
1116 461 P
1121 463 P
1246 487 P
1374 518 P
1499 557 P
1504 559 P
1629 587 P
1757 620 P
1885 647 P
2012 673 P
2523 791 P
3289 961 P
480 251 M
-31 0 R
62 0 V
-62 0 R
62 0 V
97 56 R
0 8 V
-31 -8 R
62 0 V
-62 8 R
62 0 V
96 42 R
0 8 V
-31 -8 R
62 0 V
-62 8 R
62 0 V
97 24 R
0 9 V
-31 -9 R
62 0 V
-62 9 R
62 0 V
97 24 R
0 9 V
-31 -9 R
62 0 V
-62 9 R
62 0 V
94 26 R
0 9 V
-31 -9 R
62 0 V
-62 9 R
62 0 V
-26 -13 R
0 20 V
-31 -20 R
62 0 V
-62 20 R
62 0 V
94 4 R
0 20 V
-31 -20 R
62 0 V
-62 20 R
62 0 V
97 11 R
0 20 V
-31 -20 R
62 0 V
-62 20 R
62 0 V
94 19 R
0 20 V
-31 -20 R
62 0 V
-62 20 R
62 0 V
-26 -23 R
0 30 V
-31 -30 R
62 0 V
-62 30 R
62 0 V
94 -5 R
0 36 V
-31 -36 R
62 0 V
-62 36 R
62 0 V
97 -1 R
0 33 V
-31 -33 R
62 0 V
-62 33 R
62 0 V
97 -7 R
0 33 V
-31 -33 R
62 0 V
-62 33 R
62 0 V
96 -9 R
0 37 V
-31 -37 R
62 0 V
-62 37 R
62 0 V
480 53 R
0 93 V
-31 -93 R
62 0 V
-62 93 R
62 0 V
735 49 R
0 149 V
3258 886 M
62 0 V
-62 149 R
62 0 V
LT0
480 251 P
608 401 P
735 512 P
863 597 P
991 665 P
1116 723 P
1121 724 P
1246 786 P
1374 848 P
1499 901 P
1504 894 P
1629 951 P
1757 1026 P
1885 1078 P
2012 1116 P
2523 1355 P
3289 1701 P
480 251 M
-31 0 R
62 0 V
-62 0 R
62 0 V
97 146 R
0 9 V
-31 -9 R
62 0 V
-62 9 R
62 0 V
96 102 R
0 9 V
-31 -9 R
62 0 V
-62 9 R
62 0 V
97 76 R
0 9 V
-31 -9 R
62 0 V
-62 9 R
62 0 V
97 59 R
0 9 V
-31 -9 R
62 0 V
-62 9 R
62 0 V
94 48 R
0 10 V
-31 -10 R
62 0 V
-62 10 R
62 0 V
-26 -12 R
0 16 V
-31 -16 R
62 0 V
-62 16 R
62 0 V
94 40 R
0 28 V
-31 -28 R
62 0 V
-62 28 R
62 0 V
97 40 R
0 16 V
-31 -16 R
62 0 V
-62 16 R
62 0 V
94 28 R
0 35 V
-31 -35 R
62 0 V
-62 35 R
62 0 V
-26 -45 R
0 40 V
-31 -40 R
62 0 V
-62 40 R
62 0 V
94 18 R
0 37 V
-31 -37 R
62 0 V
-62 37 R
62 0 V
97 38 R
0 38 V
-31 -38 R
62 0 V
-62 38 R
62 0 V
97 10 R
0 45 V
-31 -45 R
62 0 V
-62 45 R
62 0 V
96 -8 R
0 49 V
-31 -49 R
62 0 V
-62 49 R
62 0 V
480 167 R
0 93 V
-31 -93 R
62 0 V
-62 93 R
62 0 V
735 226 R
0 148 V
-31 -148 R
62 0 V
-62 148 R
62 0 V
LT0
1051 1923 M
180 0 V
480 251 M
128 60 V
127 50 V
128 33 V
128 33 V
125 34 V
5 2 V
125 24 V
128 31 V
125 39 V
5 2 V
125 28 V
128 33 V
128 27 V
127 26 V
511 118 V
766 170 V
LT1
1051 1823 M
180 0 V
480 251 M
608 401 L
735 512 L
128 85 V
128 68 V
125 58 V
5 1 V
125 62 V
128 62 V
125 53 V
5 -7 V
125 57 V
128 75 V
128 52 V
127 38 V
511 239 V
766 346 V
stroke
grestore
end
showpage
}
\put(991,1823){\makebox(0,0)[r]{$U_{double}$}}
\put(991,1923){\makebox(0,0)[r]{U}}
\put(1948,51){\makebox(0,0){d [a]}}
\put(3034,151){\makebox(0,0){20}}
\put(2395,151){\makebox(0,0){15}}
\put(1757,151){\makebox(0,0){10}}
\put(1118,151){\makebox(0,0){5}}
\put(480,151){\makebox(0,0){0}}
\put(420,2109){\makebox(0,0)[r]{20}}
\put(420,1645){\makebox(0,0)[r]{15}}
\put(420,1180){\makebox(0,0)[r]{10}}
\put(420,716){\makebox(0,0)[r]{5}}
\put(420,251){\makebox(0,0)[r]{0}}
\end{picture}}
\vspace{1cm}
\centerline{$\beta=1.10$}
\centerline{\input{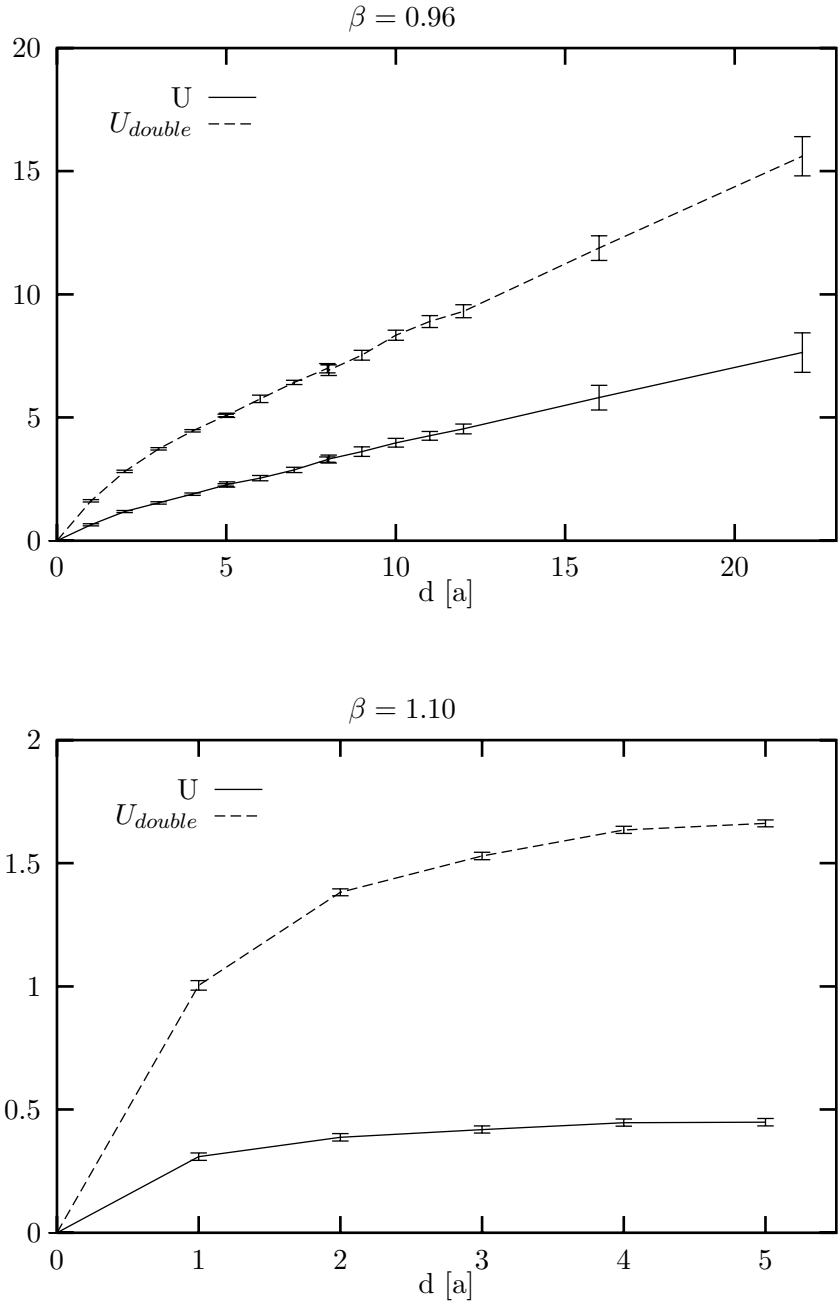}}
\caption{\label{u}Total field energy of a pair of charges (solid line) and a pair of double charges (dashed line) as a function of charge distance $d$ for $\beta=0.96$ and $\beta=1.10$.}
\end{figure}

\begin{figure}
\centerline{
\setlength{\unitlength}{0.1bp}
\special{!
/gnudict 40 dict def
gnudict begin
/Color false def
/Solid false def
/gnulinewidth 5.000 def
/vshift -33 def
/dl {10 mul} def
/hpt 31.5 def
/vpt 31.5 def
/M {moveto} bind def
/L {lineto} bind def
/R {rmoveto} bind def
/V {rlineto} bind def
/vpt2 vpt 2 mul def
/hpt2 hpt 2 mul def
/Lshow { currentpoint stroke M
  0 vshift R show } def
/Rshow { currentpoint stroke M
  dup stringwidth pop neg vshift R show } def
/Cshow { currentpoint stroke M
  dup stringwidth pop -2 div vshift R show } def
/DL { Color {setrgbcolor Solid {pop []} if 0 setdash }
 {pop pop pop Solid {pop []} if 0 setdash} ifelse } def
/BL { stroke gnulinewidth 2 mul setlinewidth } def
/AL { stroke gnulinewidth 2 div setlinewidth } def
/PL { stroke gnulinewidth setlinewidth } def
/LTb { BL [] 0 0 0 DL } def
/LTa { AL [1 dl 2 dl] 0 setdash 0 0 0 setrgbcolor } def
/LT0 { PL [] 0 1 0 DL } def
/LT1 { PL [4 dl 2 dl] 0 0 1 DL } def
/LT2 { PL [2 dl 3 dl] 1 0 0 DL } def
/LT3 { PL [1 dl 1.5 dl] 1 0 1 DL } def
/LT4 { PL [5 dl 2 dl 1 dl 2 dl] 0 1 1 DL } def
/LT5 { PL [4 dl 3 dl 1 dl 3 dl] 1 1 0 DL } def
/LT6 { PL [2 dl 2 dl 2 dl 4 dl] 0 0 0 DL } def
/LT7 { PL [2 dl 2 dl 2 dl 2 dl 2 dl 4 dl] 1 0.3 0 DL } def
/LT8 { PL [2 dl 2 dl 2 dl 2 dl 2 dl 2 dl 2 dl 4 dl] 0.5 0.5 0.5 DL } def
/P { stroke [] 0 setdash
  currentlinewidth 2 div sub M
  0 currentlinewidth V stroke } def
/D { stroke [] 0 setdash 2 copy vpt add M
  hpt neg vpt neg V hpt vpt neg V
  hpt vpt V hpt neg vpt V closepath stroke
  P } def
/A { stroke [] 0 setdash vpt sub M 0 vpt2 V
  currentpoint stroke M
  hpt neg vpt neg R hpt2 0 V stroke
  } def
/B { stroke [] 0 setdash 2 copy exch hpt sub exch vpt add M
  0 vpt2 neg V hpt2 0 V 0 vpt2 V
  hpt2 neg 0 V closepath stroke
  P } def
/C { stroke [] 0 setdash exch hpt sub exch vpt add M
  hpt2 vpt2 neg V currentpoint stroke M
  hpt2 neg 0 R hpt2 vpt2 V stroke } def
/T { stroke [] 0 setdash 2 copy vpt 1.12 mul add M
  hpt neg vpt -1.62 mul V
  hpt 2 mul 0 V
  hpt neg vpt 1.62 mul V closepath stroke
  P  } def
/S { 2 copy A C} def
end
}
\begin{picture}(3600,2160)(0,0)
\special{"
gnudict begin
gsave
50 50 translate
0.100 0.100 scale
0 setgray
/Helvetica findfont 100 scalefont setfont
newpath
-500.000000 -500.000000 translate
LTa
480 251 M
0 1858 V
LTb
480 251 M
63 0 V
2874 0 R
-63 0 V
480 716 M
63 0 V
2874 0 R
-63 0 V
480 1180 M
63 0 V
2874 0 R
-63 0 V
480 1645 M
63 0 V
2874 0 R
-63 0 V
480 2109 M
63 0 V
2874 0 R
-63 0 V
480 251 M
0 63 V
0 1795 R
0 -63 V
1067 251 M
0 63 V
0 1795 R
0 -63 V
1655 251 M
0 63 V
0 1795 R
0 -63 V
2242 251 M
0 63 V
0 1795 R
0 -63 V
2830 251 M
0 63 V
0 1795 R
0 -63 V
3417 251 M
0 63 V
0 1795 R
0 -63 V
480 251 M
2937 0 V
0 1858 V
-2937 0 V
480 251 L
LT0
3174 1946 D
597 952 D
715 887 D
832 914 D
950 882 D
1065 830 D
1070 823 D
1185 840 D
1302 824 D
1417 774 D
1422 757 D
1537 754 D
1655 766 D
1772 757 D
1890 740 D
2360 736 D
3065 736 D
3114 1946 M
180 0 V
-180 31 R
0 -62 V
180 62 R
0 -62 V
597 848 M
0 208 V
566 848 M
62 0 V
-62 208 R
62 0 V
715 828 M
0 117 V
684 828 M
62 0 V
684 945 M
62 0 V
86 -80 R
0 99 V
801 865 M
62 0 V
-62 99 R
62 0 V
950 842 M
0 81 V
919 842 M
62 0 V
-62 81 R
62 0 V
84 -126 R
0 66 V
-31 -66 R
62 0 V
-62 66 R
62 0 V
1070 755 M
0 135 V
1039 755 M
62 0 V
-62 135 R
62 0 V
84 -121 R
0 143 V
1154 769 M
62 0 V
-62 143 R
62 0 V
86 -142 R
0 107 V
1271 770 M
62 0 V
-62 107 R
62 0 V
84 -162 R
0 117 V
1386 715 M
62 0 V
-62 117 R
62 0 V
1422 680 M
0 154 V
1391 680 M
62 0 V
-62 154 R
62 0 V
84 -157 R
0 155 V
1506 677 M
62 0 V
-62 155 R
62 0 V
87 -140 R
0 148 V
1624 692 M
62 0 V
-62 148 R
62 0 V
86 -156 R
0 147 V
1741 684 M
62 0 V
-62 147 R
62 0 V
87 -165 R
0 148 V
1859 666 M
62 0 V
-62 148 R
62 0 V
2360 615 M
0 243 V
2329 615 M
62 0 V
-62 243 R
62 0 V
3065 588 M
0 296 V
3034 588 M
62 0 V
-62 296 R
62 0 V
LT1
480 716 M
30 0 V
29 0 V
30 0 V
30 0 V
29 0 V
30 0 V
30 0 V
29 0 V
30 0 V
30 0 V
29 0 V
30 0 V
30 0 V
29 0 V
30 0 V
30 0 V
29 0 V
30 0 V
30 0 V
29 0 V
30 0 V
30 0 V
29 0 V
30 0 V
30 0 V
29 0 V
30 0 V
30 0 V
29 0 V
30 0 V
30 0 V
29 0 V
30 0 V
30 0 V
29 0 V
30 0 V
30 0 V
29 0 V
30 0 V
30 0 V
29 0 V
30 0 V
30 0 V
29 0 V
30 0 V
30 0 V
29 0 V
30 0 V
30 0 V
29 0 V
30 0 V
30 0 V
29 0 V
30 0 V
30 0 V
29 0 V
30 0 V
30 0 V
29 0 V
30 0 V
30 0 V
29 0 V
30 0 V
30 0 V
29 0 V
30 0 V
30 0 V
29 0 V
30 0 V
30 0 V
29 0 V
30 0 V
30 0 V
29 0 V
30 0 V
30 0 V
29 0 V
30 0 V
30 0 V
29 0 V
30 0 V
30 0 V
29 0 V
30 0 V
30 0 V
29 0 V
30 0 V
30 0 V
29 0 V
30 0 V
30 0 V
29 0 V
30 0 V
30 0 V
29 0 V
30 0 V
30 0 V
29 0 V
30 0 V
480 1645 M
30 0 V
29 0 V
30 0 V
30 0 V
29 0 V
30 0 V
30 0 V
29 0 V
30 0 V
30 0 V
29 0 V
30 0 V
30 0 V
29 0 V
30 0 V
30 0 V
29 0 V
30 0 V
30 0 V
29 0 V
30 0 V
30 0 V
29 0 V
30 0 V
30 0 V
29 0 V
30 0 V
30 0 V
29 0 V
30 0 V
30 0 V
29 0 V
30 0 V
30 0 V
29 0 V
30 0 V
30 0 V
29 0 V
30 0 V
30 0 V
29 0 V
30 0 V
30 0 V
29 0 V
30 0 V
30 0 V
29 0 V
30 0 V
30 0 V
29 0 V
30 0 V
30 0 V
29 0 V
30 0 V
30 0 V
29 0 V
30 0 V
30 0 V
29 0 V
30 0 V
30 0 V
29 0 V
30 0 V
30 0 V
29 0 V
30 0 V
30 0 V
29 0 V
30 0 V
30 0 V
29 0 V
30 0 V
30 0 V
29 0 V
30 0 V
30 0 V
29 0 V
30 0 V
30 0 V
29 0 V
30 0 V
30 0 V
29 0 V
30 0 V
30 0 V
29 0 V
30 0 V
30 0 V
29 0 V
30 0 V
30 0 V
29 0 V
30 0 V
30 0 V
29 0 V
30 0 V
30 0 V
29 0 V
30 0 V
LT0
3174 1846 B
597 1298 B
715 1443 B
832 1482 B
950 1485 B
1067 1506 B
1185 1695 B
1302 1902 B
1420 1833 B
3114 1846 M
180 0 V
-180 31 R
0 -62 V
180 62 R
0 -62 V
597 1195 M
0 205 V
566 1195 M
62 0 V
-62 205 R
62 0 V
87 -37 R
0 161 V
684 1363 M
62 0 V
-62 161 R
62 0 V
86 -117 R
0 150 V
801 1407 M
62 0 V
-62 150 R
62 0 V
87 -143 R
0 143 V
919 1414 M
62 0 V
-62 143 R
62 0 V
86 -125 R
0 148 V
-31 -148 R
62 0 V
-62 148 R
62 0 V
87 -157 R
0 544 V
-31 -544 R
62 0 V
-62 544 R
62 0 V
86 -400 R
0 542 V
-31 -542 R
62 0 V
-62 542 R
62 0 V
87 -593 R
0 593 V
-31 -593 R
62 0 V
-62 593 R
62 0 V
stroke
grestore
end
showpage
}
\put(3054,1846){\makebox(0,0)[r]{$U_{double}/U$ ($\beta=1.10$)}}
\put(3054,1946){\makebox(0,0)[r]{$U_{double}/U$ ($\beta=0.96$)}}
\put(1948,51){\makebox(0,0){d [a]}}
\put(3417,151){\makebox(0,0){25}}
\put(2830,151){\makebox(0,0){20}}
\put(2242,151){\makebox(0,0){15}}
\put(1655,151){\makebox(0,0){10}}
\put(1067,151){\makebox(0,0){5}}
\put(480,151){\makebox(0,0){0}}
\put(420,2109){\makebox(0,0)[r]{5}}
\put(420,1645){\makebox(0,0)[r]{4}}
\put(420,1180){\makebox(0,0)[r]{3}}
\put(420,716){\makebox(0,0)[r]{2}}
\put(420,251){\makebox(0,0)[r]{1}}
\end{picture}}
\caption{\label{ratio_nd} Ratio of the total field energies $U_{double}/U$ as a function of charge distance $d$ for $\beta=0.96$ and $\beta=1.10$. In the Coulomb phase ($\beta=1.10$) this ratio is difficult to determine for larger lattice sizes.}
\end{figure}
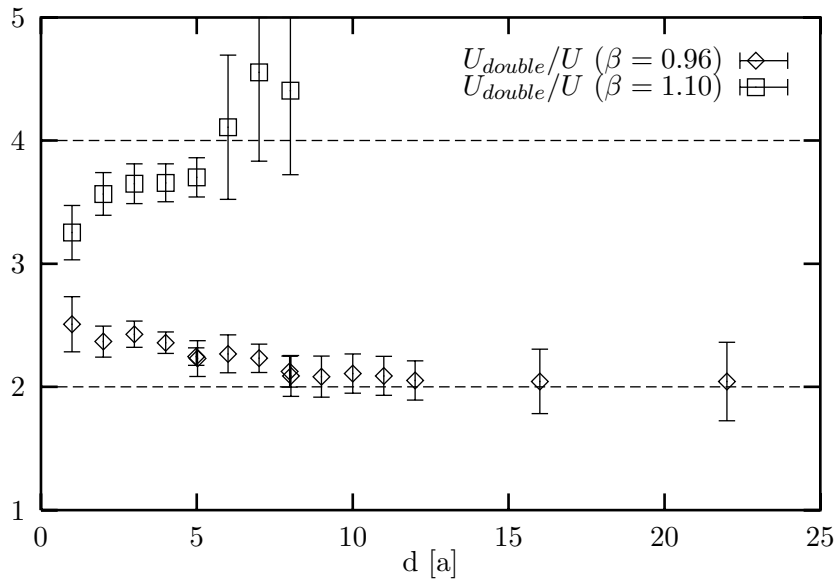

Nevertheless it is interesting to look at the total energy $U$ of the electromagnetic field as a function of charge distance. This quantity is obtained by integrating the energy density \mbox{$(E^2+B^2)/2$} over the whole lattice\footnote{We use the Minkowski space notation here. The magnetic energy density relative to the vacuum is negative, therefore it comes to cancellations between electric and magnetic energy.}, i.~e.~summing up all plaquette contributions. The significant difference of the results between confinement and Coulomb phase is demonstrated in fig.~\ref{u}, where the total field energy $U$ as a function of charge distance $d$ is plotted for single and for double flux tubes. In the upper plot ($\beta=0.96$) it can be seen that after a few lattice spacings the energy increases linearly, the ratio of slopes approximately equals 2. The lower plot shows the corresponding results at  $\beta=1.10$, where the system is in the Coulomb phase and may be approximated by classical electrodynamics. It is not possible to detect a linear increase, the ratio of energies is significantly larger (for $\beta \to \infty$ it is expected to approach 4). Simulations for this value of $\beta$ have only been performed on smaller lattices, because quantum fluctuations are much larger and the energy nearly saturates at $d \approx 5a$. The ratio of the energies of a double and a single flux tube is shown in fig.~\ref{ratio_nd}, again as a function of the charge distance $d$, for $\beta=0.96$ and $\beta=1.10$.

\section{The energy of periodically closed flux tubes}

As already pointed out, by studying closed flux tubes one gets rid of end effects which makes it easier to determine the string tension. Instead of measuring the slope of the potential at large distances, we just have to measure the energy of the flux tube relative to the vacuum\footnote{Strictly spoken the energy per unit length is not equal to the string tension because of string fluctuation contributions \cite{torelon}; this is not relevant, however, for the following discussion.}. As we will show this is a good quantity to distinguish between confined and deconfined phase. Further we want to work out the difference between the total field energy as calculated in the last section and the free energy calculated via a numerical integration over $\beta$.

\begin{figure}
\centerline{\makebox[7cm]{$3D \; U(1)$} \hspace{1cm} \makebox[7cm]{$4D \; U(1)$}}
\centerline{\epsfxsize=7cm\epsfbox{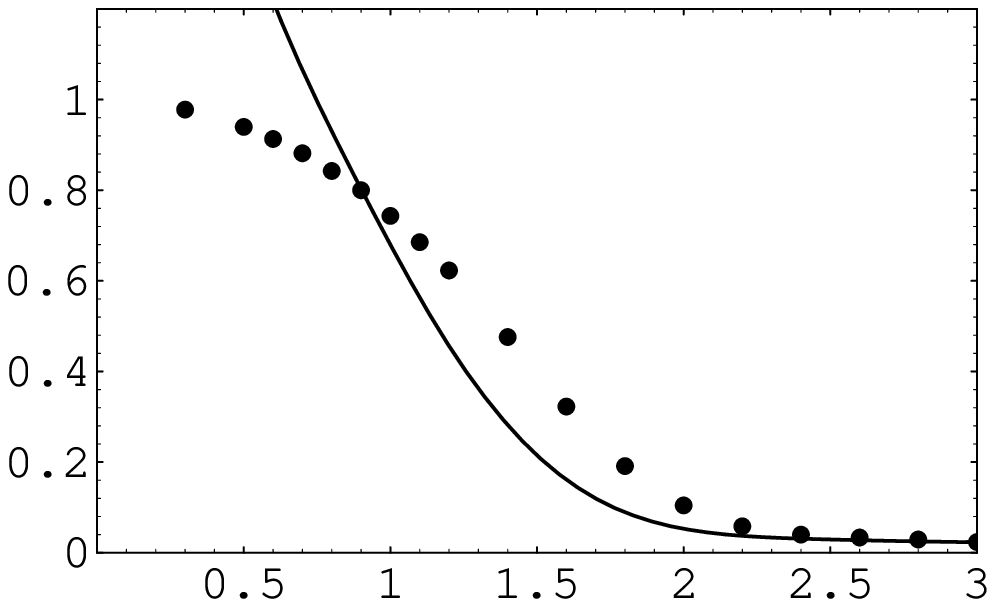}\hspace{1cm}\epsfxsize=7cm\epsfbox{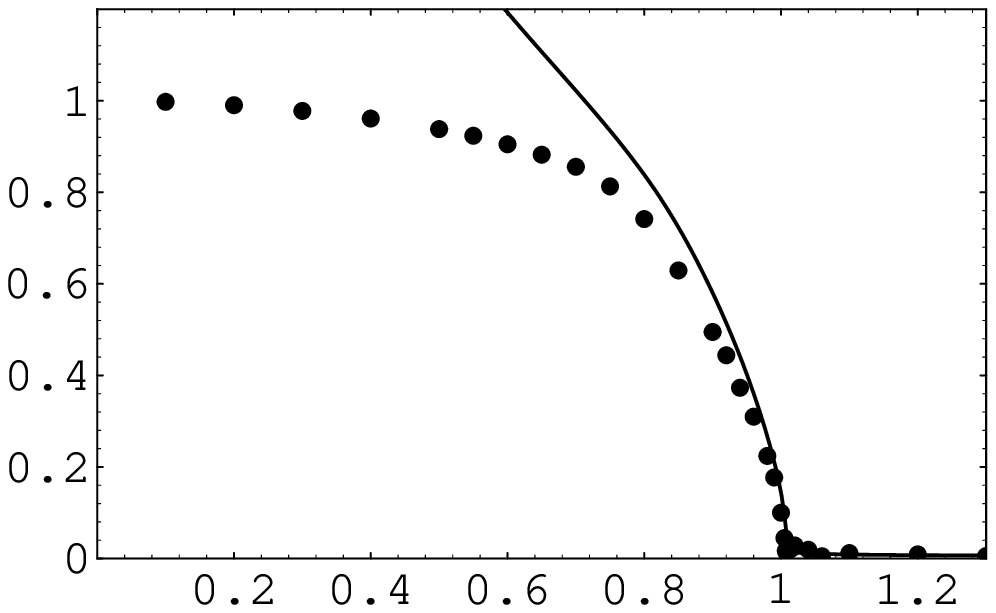}}
\centerline{\makebox[7cm]{$\beta$} \hspace{1cm} \makebox[7cm]{$\beta$}}
\caption{\label{uf}Free energy (solid line, obtained by numerical integration over $\beta$) and total field energy (points) per unit length of periodically closed flux tubes as a function of the coupling $\beta$. The left plot is for three-dimensional $U(1)$ on a $8^2 \times 16$ lattice, the right plot for the four-dimensional theory on a $8^3 \times 16$ lattice. Errorbars are not drawn, since the estimated relative error for the free energy does not exceed a few percent, even in the critical $\beta$-region of the phase transition.}
\end{figure}

In fig.~\ref{uf} free energy and field energy per length are contrasted both for the three- and for the four-dimensional theory as a function of $\beta$. A large difference between the quantities is observed at strong coupling. For $\beta \to 0$ the free energy diverges logarithmically (in both three and four dimensions), whereas the field energy per length tends to 1 (in units of the inverse squared lattice spacing). In the classical limit, however, both energies show the same behaviour.

In three dimensions the free energy becomes smaller than the field energy at $\beta \approx 0.9$, as can be seen from fig.~\ref{uf}. In a range of weaker couplings where already many simulations have been performed in previous times, the discrepancy is approximately a factor 2 which is in agreement with ref.~\cite{greensite}. In the four-dimensional theory the free energy is greater than the field energy for the whole confinement phase. At the phase transition both of them show a drastic decrease. On a lattice of finite transverse extent the energy is not exactly zero above the phase transition due to the classical energy of the corresponding homogeneous electric field on a torus. This kind of finite size effect may play a more important role in three dimensions as will be discussed in the next section. Thus we have seen that it is necessary to make a difference between free energy and total field energy, nevertheless these two lattice quantities can be related to each other by sum rules \cite{michael} which may be used to extract further informations on the system \cite{petrus}. Corresponding calculations are in progress and will be reported elsewhere.

\begin{figure}
\centerline{\makebox[10cm]{$F_{double}/F$}}
\centerline{\epsfxsize=10cm\epsfbox{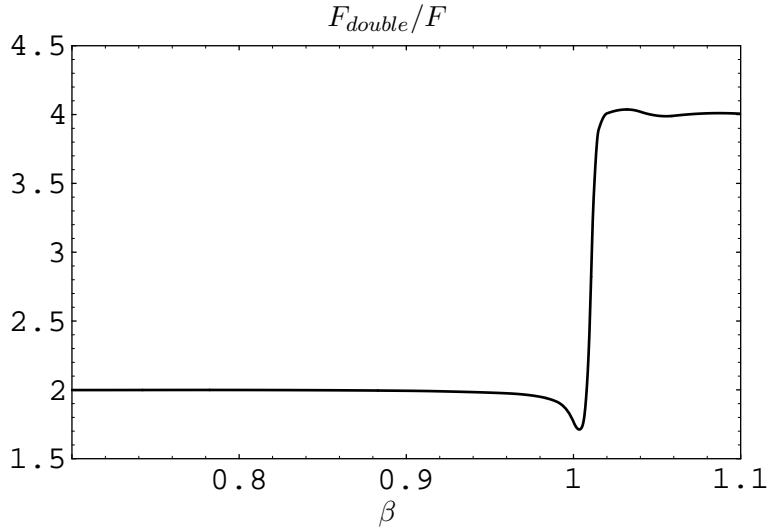}}
\centerline{\makebox[10cm]{$\beta$}}
\caption{\label{ratio}Ratio of free energies of doubly and singly charged closed flux tubes as a function of $\beta$ for four-dimensional $U(1)$ on a $8^3 \times 16$ lattice.}
\end{figure}

Now we proceed to the investigation of closed double flux tubes in order to continue the discussion of the last section. For the moment we focus on the four-dimensional theory; differences to the three-dimensional case will be discussed later. The free energy of a double flux tube shows a very similar $\beta$-dependence as for the single flux tube. We are mostly interested in the ratio $F_{double}/F$; it is plotted in fig.~\ref{ratio} as a function of the coupling $\beta$. 
In the confinement phase (starting from very small $\beta$-values not shown in the plot) the ratio is 2 within the estimated errors, which could have been expected from the results of the last section. It can be also noticed, however, that between $\beta \approx 0.9$ and the phase transition at $\beta \approx 1.01$ this ratio drops significantly below 2. Above the phase transition, in the Coulomb phase, the ratio jumps to 4 which is the classically expected result for the energy of a homogeneous field of double strength. The results shown in fig.~\ref{ratio} are for a $8^3 \times 16$ lattice; it has also been checked that there does not change much when considering larger spatial extents except that the statistics gets worse.

How to understand these results in terms of the dual formulation? That the energy of a double flux tube is just twice the energy of a single flux tube in the strong coupling limit has a simple explanation. Let us start from a configuration with the electric flux identical to the double dual Dirac sheet ($^*n=2 \Rightarrow {}^*k=2$). The action of such doubly occupied plaquettes is higher than that of two singly occupied plaquettes for any finite value of $\beta$. Therefore it is favourable for the system to form two completely separated flux strings (this is accomplished by a change in $^*l$) which in the ground state of the dual theory do not interact at all\footnote{It is not correct, however, to conclude from this ``classical'' behaviour of the dual theory that the confinement vacuum behaves like a dual type-II superconductor. As is shown in fig.~\ref{ratio} and will be analyzed later, the quantum fluctuations of the dual theory can lead to an attraction between the $U(1)$ flux tubes.}. For short open strings between doubly charged static sources, end effects cause a deviation from the ratio two.

In ref.~\cite{trottier} it was argued that one should expect that the string tension scales like the squared charge through an extension of the Villain approximation used in ref.~\cite{banks}, including multiply charged Wilson loops. This argument is only true, however, if the flux strings cover the same plaquette. As we explained, such configurations are unstable in the dual simulation. An analysis of the results obtained in ref.~\cite{trottier} for the three-dimensional theory will be performed in the next section.

On the other hand, the classical Coulomb energy of a doubly charged state is four times higher. This can also be explained within the dual formulation. In the Coulomb phase the probability for higher plaquette values $^*k$ increases; for large $\beta$ the $k$-distribution may be approximated by a continuous function $\rho(k)$ which is symmetric with respect to $k=0$ if no charges are present. By introducing a closed single flux state on the lattice, this distribution is shifted a little by $\delta k_0$ so that $\rho(k) \to \rho(k-\delta k_0)$; two flux quanta cause a shift by $2 \delta k_0$. The operator for the squared field strength introduced in eq.~(\ref{cos}) in the weak coupling limit behaves as $k^2/2\beta + \mathit{const}(\beta)$. In this case the total energy of a flux state with respect to the vacuum is proportional to
\begin{equation}
\int dk \; \left[ \rho(k-\delta k_0) - \rho(k) \right] \; \frac{k^2}{2\beta} \;\; = \;\; \int dk \; \rho(k) \left[ \frac{(k+\delta k_0)^2}{2\beta} - \frac{k^2}{2\beta} \right] \;\; \propto \;\; (\delta k_0)^2,
\end{equation}
i.~e.~it scales with the squared charge of the system.

A physically very interesting observation is the relative decrease of the double flux tube energy in the region of $\beta$ approaching the phase transition. This indicates that there should be an attractive force between flux tubes in four dimensions. A closer investigation and discussion of this effect will be performed in section \ref{ia}.

\begin{figure}
\centerline{
\setlength{\unitlength}{0.1bp}
\special{!
/gnudict 40 dict def
gnudict begin
/Color false def
/Solid false def
/gnulinewidth 5.000 def
/vshift -33 def
/dl {10 mul} def
/hpt 31.5 def
/vpt 31.5 def
/M {moveto} bind def
/L {lineto} bind def
/R {rmoveto} bind def
/V {rlineto} bind def
/vpt2 vpt 2 mul def
/hpt2 hpt 2 mul def
/Lshow { currentpoint stroke M
  0 vshift R show } def
/Rshow { currentpoint stroke M
  dup stringwidth pop neg vshift R show } def
/Cshow { currentpoint stroke M
  dup stringwidth pop -2 div vshift R show } def
/DL { Color {setrgbcolor Solid {pop []} if 0 setdash }
 {pop pop pop Solid {pop []} if 0 setdash} ifelse } def
/BL { stroke gnulinewidth 2 mul setlinewidth } def
/AL { stroke gnulinewidth 2 div setlinewidth } def
/PL { stroke gnulinewidth setlinewidth } def
/LTb { BL [] 0 0 0 DL } def
/LTa { AL [1 dl 2 dl] 0 setdash 0 0 0 setrgbcolor } def
/LT0 { PL [] 0 1 0 DL } def
/LT1 { PL [4 dl 2 dl] 0 0 1 DL } def
/LT2 { PL [2 dl 3 dl] 1 0 0 DL } def
/LT3 { PL [1 dl 1.5 dl] 1 0 1 DL } def
/LT4 { PL [5 dl 2 dl 1 dl 2 dl] 0 1 1 DL } def
/LT5 { PL [4 dl 3 dl 1 dl 3 dl] 1 1 0 DL } def
/LT6 { PL [2 dl 2 dl 2 dl 4 dl] 0 0 0 DL } def
/LT7 { PL [2 dl 2 dl 2 dl 2 dl 2 dl 4 dl] 1 0.3 0 DL } def
/LT8 { PL [2 dl 2 dl 2 dl 2 dl 2 dl 2 dl 2 dl 4 dl] 0.5 0.5 0.5 DL } def
/P { stroke [] 0 setdash
  currentlinewidth 2 div sub M
  0 currentlinewidth V stroke } def
/D { stroke [] 0 setdash 2 copy vpt add M
  hpt neg vpt neg V hpt vpt neg V
  hpt vpt V hpt neg vpt V closepath stroke
  P } def
/A { stroke [] 0 setdash vpt sub M 0 vpt2 V
  currentpoint stroke M
  hpt neg vpt neg R hpt2 0 V stroke
  } def
/B { stroke [] 0 setdash 2 copy exch hpt sub exch vpt add M
  0 vpt2 neg V hpt2 0 V 0 vpt2 V
  hpt2 neg 0 V closepath stroke
  P } def
/C { stroke [] 0 setdash exch hpt sub exch vpt add M
  hpt2 vpt2 neg V currentpoint stroke M
  hpt2 neg 0 R hpt2 vpt2 V stroke } def
/T { stroke [] 0 setdash 2 copy vpt 1.12 mul add M
  hpt neg vpt -1.62 mul V
  hpt 2 mul 0 V
  hpt neg vpt 1.62 mul V closepath stroke
  P  } def
/S { 2 copy A C} def
end
}
\begin{picture}(3600,2160)(0,0)
\special{"
gnudict begin
gsave
50 50 translate
0.100 0.100 scale
0 setgray
/Helvetica findfont 100 scalefont setfont
newpath
-500.000000 -500.000000 translate
LTa
LTb
480 251 M
63 0 V
2874 0 R
-63 0 V
480 623 M
63 0 V
2874 0 R
-63 0 V
480 994 M
63 0 V
2874 0 R
-63 0 V
480 1366 M
63 0 V
2874 0 R
-63 0 V
480 1737 M
63 0 V
2874 0 R
-63 0 V
480 2109 M
63 0 V
2874 0 R
-63 0 V
480 251 M
0 63 V
0 1795 R
0 -63 V
954 251 M
0 63 V
0 1795 R
0 -63 V
1427 251 M
0 63 V
0 1795 R
0 -63 V
1901 251 M
0 63 V
0 1795 R
0 -63 V
2375 251 M
0 63 V
0 1795 R
0 -63 V
2849 251 M
0 63 V
0 1795 R
0 -63 V
3322 251 M
0 63 V
0 1795 R
0 -63 V
480 251 M
2937 0 V
0 1858 V
-2937 0 V
480 251 L
LT0
1311 1923 D
480 623 D
717 623 D
954 623 D
1191 625 D
1427 622 D
1664 622 D
1901 621 D
2138 621 D
2375 615 D
2470 602 D
2564 602 D
2659 579 D
2754 575 D
2801 519 D
2849 403 D
2872 700 D
2896 1748 D
2920 1318 D
2943 1125 D
3038 1211 D
3322 1686 D
1251 1923 M
180 0 V
-180 31 R
0 -62 V
180 62 R
0 -62 V
480 622 M
0 2 V
-31 -2 R
62 0 V
-62 2 R
62 0 V
206 -2 R
0 2 V
-31 -2 R
62 0 V
-62 2 R
62 0 V
206 -2 R
0 3 V
-31 -3 R
62 0 V
-62 3 R
62 0 V
206 -1 R
0 3 V
-31 -3 R
62 0 V
-62 3 R
62 0 V
205 -8 R
0 5 V
-31 -5 R
62 0 V
-62 5 R
62 0 V
206 -5 R
0 6 V
-31 -6 R
62 0 V
-62 6 R
62 0 V
206 -8 R
0 8 V
-31 -8 R
62 0 V
-62 8 R
62 0 V
206 -9 R
0 10 V
-31 -10 R
62 0 V
-62 10 R
62 0 V
206 -19 R
0 16 V
-31 -16 R
62 0 V
-62 16 R
62 0 V
64 -30 R
0 18 V
-31 -18 R
62 0 V
-62 18 R
62 0 V
63 -21 R
0 23 V
-31 -23 R
62 0 V
-62 23 R
62 0 V
64 -49 R
0 30 V
-31 -30 R
62 0 V
-62 30 R
62 0 V
64 -41 R
0 43 V
-31 -43 R
62 0 V
-62 43 R
62 0 V
16 -104 R
0 54 V
-31 -54 R
62 0 V
-62 54 R
62 0 V
17 -187 R
0 87 V
-31 -87 R
62 0 V
-62 87 R
62 0 V
-8 126 R
0 257 V
2841 572 M
62 0 V
-62 257 R
62 0 V
-7 153 R
0 1127 V
2865 982 M
62 0 V
-62 1127 R
62 0 V
2920 821 M
0 994 V
2889 821 M
62 0 V
-62 994 R
62 0 V
2943 802 M
0 646 V
2912 802 M
62 0 V
-62 646 R
62 0 V
64 -718 R
0 961 V
3007 730 M
62 0 V
-62 961 R
62 0 V
3322 707 M
0 1402 V
3291 707 M
62 0 V
-62 1402 R
62 0 V
LT2
480 623 M
30 0 V
29 0 V
30 0 V
30 0 V
29 0 V
30 0 V
30 0 V
29 0 V
30 0 V
30 0 V
29 0 V
30 0 V
30 0 V
29 0 V
30 0 V
30 0 V
29 0 V
30 0 V
30 0 V
29 0 V
30 0 V
30 0 V
29 0 V
30 0 V
30 0 V
29 0 V
30 0 V
30 0 V
29 0 V
30 0 V
30 0 V
29 0 V
30 0 V
30 0 V
29 0 V
30 0 V
30 0 V
29 0 V
30 0 V
30 0 V
29 0 V
30 0 V
30 0 V
29 0 V
30 0 V
30 0 V
29 0 V
30 0 V
30 0 V
29 0 V
30 0 V
30 0 V
29 0 V
30 0 V
30 0 V
29 0 V
30 0 V
30 0 V
29 0 V
30 0 V
30 0 V
29 0 V
30 0 V
30 0 V
29 0 V
30 0 V
30 0 V
29 0 V
30 0 V
30 0 V
29 0 V
30 0 V
30 0 V
29 0 V
30 0 V
30 0 V
29 0 V
30 0 V
30 0 V
29 0 V
30 0 V
30 0 V
29 0 V
30 0 V
30 0 V
29 0 V
30 0 V
30 0 V
29 0 V
30 0 V
30 0 V
29 0 V
30 0 V
30 0 V
29 0 V
30 0 V
30 0 V
29 0 V
30 0 V
480 1366 M
30 0 V
29 0 V
30 0 V
30 0 V
29 0 V
30 0 V
30 0 V
29 0 V
30 0 V
30 0 V
29 0 V
30 0 V
30 0 V
29 0 V
30 0 V
30 0 V
29 0 V
30 0 V
30 0 V
29 0 V
30 0 V
30 0 V
29 0 V
30 0 V
30 0 V
29 0 V
30 0 V
30 0 V
29 0 V
30 0 V
30 0 V
29 0 V
30 0 V
30 0 V
29 0 V
30 0 V
30 0 V
29 0 V
30 0 V
30 0 V
29 0 V
30 0 V
30 0 V
29 0 V
30 0 V
30 0 V
29 0 V
30 0 V
30 0 V
29 0 V
30 0 V
30 0 V
29 0 V
30 0 V
30 0 V
29 0 V
30 0 V
30 0 V
29 0 V
30 0 V
30 0 V
29 0 V
30 0 V
30 0 V
29 0 V
30 0 V
30 0 V
29 0 V
30 0 V
30 0 V
29 0 V
30 0 V
30 0 V
29 0 V
30 0 V
30 0 V
29 0 V
30 0 V
30 0 V
29 0 V
30 0 V
30 0 V
29 0 V
30 0 V
30 0 V
29 0 V
30 0 V
30 0 V
29 0 V
30 0 V
30 0 V
29 0 V
30 0 V
30 0 V
29 0 V
30 0 V
30 0 V
29 0 V
30 0 V
stroke
grestore
end
showpage
}
\put(1191,1923){\makebox(0,0)[r]{$U_{double}/U$}}
\put(1948,51){\makebox(0,0){$\beta$}}
\put(3322,151){\makebox(0,0){1.1}}
\put(2849,151){\makebox(0,0){1}}
\put(2375,151){\makebox(0,0){0.9}}
\put(1901,151){\makebox(0,0){0.8}}
\put(1427,151){\makebox(0,0){0.7}}
\put(954,151){\makebox(0,0){0.6}}
\put(480,151){\makebox(0,0){0.5}}
\put(420,2109){\makebox(0,0)[r]{6}}
\put(420,1737){\makebox(0,0)[r]{5}}
\put(420,1366){\makebox(0,0)[r]{4}}
\put(420,994){\makebox(0,0)[r]{3}}
\put(420,623){\makebox(0,0)[r]{2}}
\put(420,251){\makebox(0,0)[r]{1}}
\end{picture}}
\caption{\label{factor_u} Ratio of total field energies of doubly and singly charged closed flux tubes as a function of $\beta$ for four-dimensional $U(1)$.}
\end{figure}
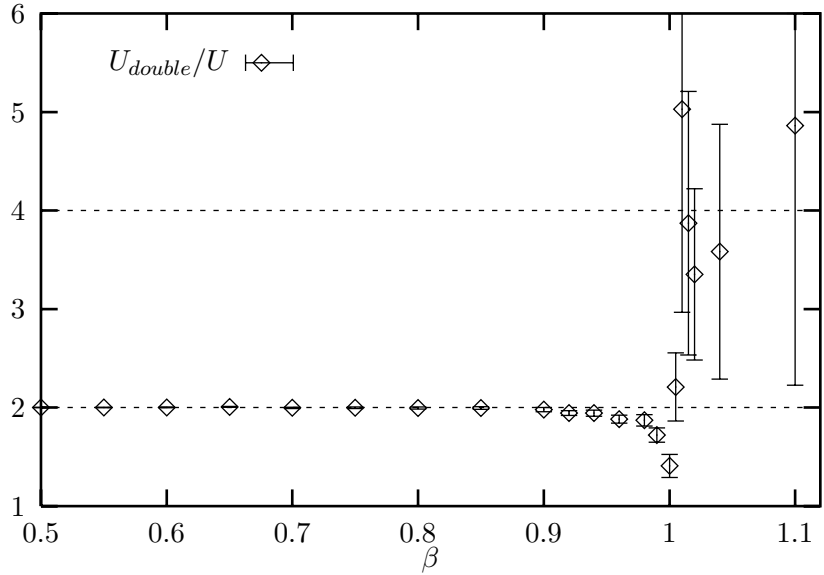

Finally we also show the above discussed ratio between double and single flux tubes for the total field energy in fig.~\ref{factor_u}. It qualitatively behaves very similarly to the ratio of free energies and also confirms the observed decrease below 2 in the confinement phase in the vicinity of the phase transition. Because the field energy of a flux state in the Coulomb phase is very small as discussed above, the errorbars for the corresponding ratios get very large.

\section{Analysis of the three-dimensional theory}

As already mentioned our results differ from those of ref.~\cite{trottier} which were obtained in a simulation of three-dimensional $U(1)$ lattice gauge theory. Singly and doubly charged Wilson loops were calculated at $\beta=2.4$ on a $32^3$ lattice, and the potential was found to scale with the squared charge within a few percent. Due to the nature of a standard calculation of correlation functions, the largest considered spatial extent of the loop was $6a$.

We will argue that the observed behaviour may be due to two effects which are relevant mainly for the three-dimensional theory: the finite transverse extent of the lattice (there is only one transverse direction for flux tubes in three dimensions) and the influence of the static sources, yielding a non-negligible Coulomb term in the potential (which is logarithmic in three dimensions) for small and intermediate charge distances.

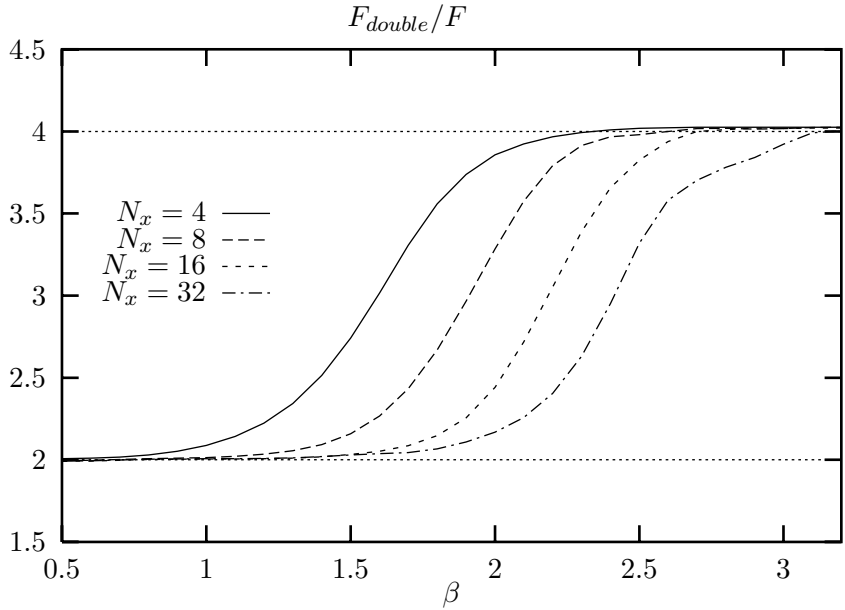
\begin{figure}
\centerline{\makebox[10cm]{$F_{double}/F$}}
\centerline{
\setlength{\unitlength}{0.1bp}
\special{!
/gnudict 40 dict def
gnudict begin
/Color false def
/Solid false def
/gnulinewidth 5.000 def
/vshift -33 def
/dl {10 mul} def
/hpt 31.5 def
/vpt 31.5 def
/M {moveto} bind def
/L {lineto} bind def
/R {rmoveto} bind def
/V {rlineto} bind def
/vpt2 vpt 2 mul def
/hpt2 hpt 2 mul def
/Lshow { currentpoint stroke M
  0 vshift R show } def
/Rshow { currentpoint stroke M
  dup stringwidth pop neg vshift R show } def
/Cshow { currentpoint stroke M
  dup stringwidth pop -2 div vshift R show } def
/DL { Color {setrgbcolor Solid {pop []} if 0 setdash }
 {pop pop pop Solid {pop []} if 0 setdash} ifelse } def
/BL { stroke gnulinewidth 2 mul setlinewidth } def
/AL { stroke gnulinewidth 2 div setlinewidth } def
/PL { stroke gnulinewidth setlinewidth } def
/LTb { BL [] 0 0 0 DL } def
/LTa { AL [1 dl 2 dl] 0 setdash 0 0 0 setrgbcolor } def
/LT0 { PL [] 0 1 0 DL } def
/LT1 { PL [4 dl 2 dl] 0 0 1 DL } def
/LT2 { PL [2 dl 3 dl] 1 0 0 DL } def
/LT3 { PL [1 dl 1.5 dl] 1 0 1 DL } def
/LT4 { PL [5 dl 2 dl 1 dl 2 dl] 0 1 1 DL } def
/LT5 { PL [4 dl 3 dl 1 dl 3 dl] 1 1 0 DL } def
/LT6 { PL [2 dl 2 dl 2 dl 4 dl] 0 0 0 DL } def
/LT7 { PL [2 dl 2 dl 2 dl 2 dl 2 dl 4 dl] 1 0.3 0 DL } def
/LT8 { PL [2 dl 2 dl 2 dl 2 dl 2 dl 2 dl 2 dl 4 dl] 0.5 0.5 0.5 DL } def
/P { stroke [] 0 setdash
  currentlinewidth 2 div sub M
  0 currentlinewidth V stroke } def
/D { stroke [] 0 setdash 2 copy vpt add M
  hpt neg vpt neg V hpt vpt neg V
  hpt vpt V hpt neg vpt V closepath stroke
  P } def
/A { stroke [] 0 setdash vpt sub M 0 vpt2 V
  currentpoint stroke M
  hpt neg vpt neg R hpt2 0 V stroke
  } def
/B { stroke [] 0 setdash 2 copy exch hpt sub exch vpt add M
  0 vpt2 neg V hpt2 0 V 0 vpt2 V
  hpt2 neg 0 V closepath stroke
  P } def
/C { stroke [] 0 setdash exch hpt sub exch vpt add M
  hpt2 vpt2 neg V currentpoint stroke M
  hpt2 neg 0 R hpt2 vpt2 V stroke } def
/T { stroke [] 0 setdash 2 copy vpt 1.12 mul add M
  hpt neg vpt -1.62 mul V
  hpt 2 mul 0 V
  hpt neg vpt 1.62 mul V closepath stroke
  P  } def
/S { 2 copy A C} def
end
}
\begin{picture}(3600,2160)(0,0)
\special{"
gnudict begin
gsave
50 50 translate
0.100 0.100 scale
0 setgray
/Helvetica findfont 100 scalefont setfont
newpath
-500.000000 -500.000000 translate
LTa
LTb
480 251 M
63 0 V
2874 0 R
-63 0 V
480 561 M
63 0 V
2874 0 R
-63 0 V
480 870 M
63 0 V
2874 0 R
-63 0 V
480 1180 M
63 0 V
2874 0 R
-63 0 V
480 1490 M
63 0 V
2874 0 R
-63 0 V
480 1799 M
63 0 V
2874 0 R
-63 0 V
480 2109 M
63 0 V
2874 0 R
-63 0 V
480 251 M
0 63 V
0 1795 R
0 -63 V
1024 251 M
0 63 V
0 1795 R
0 -63 V
1568 251 M
0 63 V
0 1795 R
0 -63 V
2112 251 M
0 63 V
0 1795 R
0 -63 V
2656 251 M
0 63 V
0 1795 R
0 -63 V
3199 251 M
0 63 V
0 1795 R
0 -63 V
480 251 M
2937 0 V
0 1858 V
-2937 0 V
480 251 L
LT0
1084 1490 M
180 0 V
480 565 M
109 2 V
109 4 V
108 8 V
109 14 V
109 22 V
109 34 V
108 50 V
109 74 V
109 106 V
109 141 V
109 170 V
108 180 V
109 156 V
109 111 V
109 74 V
108 41 V
109 27 V
109 16 V
109 10 V
109 6 V
108 2 V
109 2 V
109 0 V
109 0 V
108 0 V
109 0 V
109 0 V
LT1
1084 1390 M
180 0 V
480 558 M
109 1 V
109 3 V
108 3 V
109 2 V
109 2 V
109 5 V
108 8 V
109 13 V
109 23 V
109 41 V
109 67 V
108 103 V
109 146 V
109 184 V
109 197 V
108 180 V
109 134 V
109 76 V
109 33 V
109 8 V
108 12 V
109 11 V
109 1 V
109 -3 V
108 2 V
109 4 V
109 1 V
LT2
1084 1290 M
180 0 V
480 555 M
109 1 V
109 3 V
108 3 V
109 1 V
109 1 V
109 0 V
108 1 V
109 2 V
109 5 V
109 9 V
109 12 V
108 21 V
109 38 V
109 67 V
109 115 V
108 171 V
109 207 V
109 209 V
109 166 V
109 104 V
108 70 V
109 38 V
109 7 V
109 6 V
108 -2 V
109 2 V
109 0 V
LT4
1084 1190 M
180 0 V
480 556 M
109 1 V
109 3 V
108 3 V
109 1 V
109 1 V
109 0 V
108 1 V
109 2 V
109 5 V
109 6 V
109 5 V
108 4 V
109 14 V
109 26 V
109 37 V
108 55 V
109 91 V
109 140 V
109 199 V
109 228 V
108 162 V
109 75 V
109 49 V
109 37 V
108 50 V
109 43 V
109 7 V
LT3
480 561 M
30 0 V
29 0 V
30 0 V
30 0 V
29 0 V
30 0 V
30 0 V
29 0 V
30 0 V
30 0 V
29 0 V
30 0 V
30 0 V
29 0 V
30 0 V
30 0 V
29 0 V
30 0 V
30 0 V
29 0 V
30 0 V
30 0 V
29 0 V
30 0 V
30 0 V
29 0 V
30 0 V
30 0 V
29 0 V
30 0 V
30 0 V
29 0 V
30 0 V
30 0 V
29 0 V
30 0 V
30 0 V
29 0 V
30 0 V
30 0 V
29 0 V
30 0 V
30 0 V
29 0 V
30 0 V
30 0 V
29 0 V
30 0 V
30 0 V
29 0 V
30 0 V
30 0 V
29 0 V
30 0 V
30 0 V
29 0 V
30 0 V
30 0 V
29 0 V
30 0 V
30 0 V
29 0 V
30 0 V
30 0 V
29 0 V
30 0 V
30 0 V
29 0 V
30 0 V
30 0 V
29 0 V
30 0 V
30 0 V
29 0 V
30 0 V
30 0 V
29 0 V
30 0 V
30 0 V
29 0 V
30 0 V
30 0 V
29 0 V
30 0 V
30 0 V
29 0 V
30 0 V
30 0 V
29 0 V
30 0 V
30 0 V
29 0 V
30 0 V
30 0 V
29 0 V
30 0 V
30 0 V
29 0 V
30 0 V
480 1799 M
30 0 V
29 0 V
30 0 V
30 0 V
29 0 V
30 0 V
30 0 V
29 0 V
30 0 V
30 0 V
29 0 V
30 0 V
30 0 V
29 0 V
30 0 V
30 0 V
29 0 V
30 0 V
30 0 V
29 0 V
30 0 V
30 0 V
29 0 V
30 0 V
30 0 V
29 0 V
30 0 V
30 0 V
29 0 V
30 0 V
30 0 V
29 0 V
30 0 V
30 0 V
29 0 V
30 0 V
30 0 V
29 0 V
30 0 V
30 0 V
29 0 V
30 0 V
30 0 V
29 0 V
30 0 V
30 0 V
29 0 V
30 0 V
30 0 V
29 0 V
30 0 V
30 0 V
29 0 V
30 0 V
30 0 V
29 0 V
30 0 V
30 0 V
29 0 V
30 0 V
30 0 V
29 0 V
30 0 V
30 0 V
29 0 V
30 0 V
30 0 V
29 0 V
30 0 V
30 0 V
29 0 V
30 0 V
30 0 V
29 0 V
30 0 V
30 0 V
29 0 V
30 0 V
30 0 V
29 0 V
30 0 V
30 0 V
29 0 V
30 0 V
30 0 V
29 0 V
30 0 V
30 0 V
29 0 V
30 0 V
30 0 V
29 0 V
30 0 V
30 0 V
29 0 V
30 0 V
30 0 V
29 0 V
30 0 V
stroke
grestore
end
showpage
}
\put(1024,1190){\makebox(0,0)[r]{$N_x=32$}}
\put(1024,1290){\makebox(0,0)[r]{$N_x=16$}}
\put(1024,1390){\makebox(0,0)[r]{$N_x=8$}}
\put(1024,1490){\makebox(0,0)[r]{$N_x=4$}}
\put(1948,51){\makebox(0,0){$\beta$}}
\put(3199,151){\makebox(0,0){3}}
\put(2656,151){\makebox(0,0){2.5}}
\put(2112,151){\makebox(0,0){2}}
\put(1568,151){\makebox(0,0){1.5}}
\put(1024,151){\makebox(0,0){1}}
\put(480,151){\makebox(0,0){0.5}}
\put(420,2109){\makebox(0,0)[r]{4.5}}
\put(420,1799){\makebox(0,0)[r]{4}}
\put(420,1490){\makebox(0,0)[r]{3.5}}
\put(420,1180){\makebox(0,0)[r]{3}}
\put(420,870){\makebox(0,0)[r]{2.5}}
\put(420,561){\makebox(0,0)[r]{2}}
\put(420,251){\makebox(0,0)[r]{1.5}}
\end{picture}}
\caption{\label{ratio_3d}The ratio of the free energies of doubly and singly charged closed flux tubes as a function of $\beta$ in three-dimensional $U(1)$ for various transverse lattice sizes $N_x$.}
\end{figure}

Let us first look again at the ratio of free energies for closed flux tubes. Since in three-dimensional $U(1)$ the string tension is non-zero for all values of the coupling \cite{polyakov} we might expect that the ratio $F_{double}/F$ equals 2 for all values of $\beta$. But if $\beta$ is sufficiently large, the thickness of the flux tube approaches the transverse lattice size, and the cross section cannot increase proportionally to the charge any more. This causes a smooth transition of $F_{double}/F$ from 2 to 4. The dependence of this transition on the transverse extent $N_x$ of the lattice is illustrated in fig.~\ref{ratio_3d}. By enlarging $N_x$ the transition can be shifted to higher values of $\beta$. Consider now $\beta=2.4$ and $N_x=32$ for comparison to ref.~\cite{trottier}: We are just in the transition region, the free energy of the closed double flux tube is approximately three times higher than that of the single flux tube. This shows that finite size effects already play a considerable role. We have also investigated the ratio of total field energies; again this ratio behaves very similarly.

\begin{figure}
\centerline{
\setlength{\unitlength}{0.1bp}
\special{!
/gnudict 40 dict def
gnudict begin
/Color false def
/Solid false def
/gnulinewidth 5.000 def
/vshift -33 def
/dl {10 mul} def
/hpt 31.5 def
/vpt 31.5 def
/M {moveto} bind def
/L {lineto} bind def
/R {rmoveto} bind def
/V {rlineto} bind def
/vpt2 vpt 2 mul def
/hpt2 hpt 2 mul def
/Lshow { currentpoint stroke M
  0 vshift R show } def
/Rshow { currentpoint stroke M
  dup stringwidth pop neg vshift R show } def
/Cshow { currentpoint stroke M
  dup stringwidth pop -2 div vshift R show } def
/DL { Color {setrgbcolor Solid {pop []} if 0 setdash }
 {pop pop pop Solid {pop []} if 0 setdash} ifelse } def
/BL { stroke gnulinewidth 2 mul setlinewidth } def
/AL { stroke gnulinewidth 2 div setlinewidth } def
/PL { stroke gnulinewidth setlinewidth } def
/LTb { BL [] 0 0 0 DL } def
/LTa { AL [1 dl 2 dl] 0 setdash 0 0 0 setrgbcolor } def
/LT0 { PL [] 0 1 0 DL } def
/LT1 { PL [4 dl 2 dl] 0 0 1 DL } def
/LT2 { PL [2 dl 3 dl] 1 0 0 DL } def
/LT3 { PL [1 dl 1.5 dl] 1 0 1 DL } def
/LT4 { PL [5 dl 2 dl 1 dl 2 dl] 0 1 1 DL } def
/LT5 { PL [4 dl 3 dl 1 dl 3 dl] 1 1 0 DL } def
/LT6 { PL [2 dl 2 dl 2 dl 4 dl] 0 0 0 DL } def
/LT7 { PL [2 dl 2 dl 2 dl 2 dl 2 dl 4 dl] 1 0.3 0 DL } def
/LT8 { PL [2 dl 2 dl 2 dl 2 dl 2 dl 2 dl 2 dl 4 dl] 0.5 0.5 0.5 DL } def
/P { stroke [] 0 setdash
  currentlinewidth 2 div sub M
  0 currentlinewidth V stroke } def
/D { stroke [] 0 setdash 2 copy vpt add M
  hpt neg vpt neg V hpt vpt neg V
  hpt vpt V hpt neg vpt V closepath stroke
  P } def
/A { stroke [] 0 setdash vpt sub M 0 vpt2 V
  currentpoint stroke M
  hpt neg vpt neg R hpt2 0 V stroke
  } def
/B { stroke [] 0 setdash 2 copy exch hpt sub exch vpt add M
  0 vpt2 neg V hpt2 0 V 0 vpt2 V
  hpt2 neg 0 V closepath stroke
  P } def
/C { stroke [] 0 setdash exch hpt sub exch vpt add M
  hpt2 vpt2 neg V currentpoint stroke M
  hpt2 neg 0 R hpt2 vpt2 V stroke } def
/T { stroke [] 0 setdash 2 copy vpt 1.12 mul add M
  hpt neg vpt -1.62 mul V
  hpt 2 mul 0 V
  hpt neg vpt 1.62 mul V closepath stroke
  P  } def
/S { 2 copy A C} def
end
}
\begin{picture}(3600,2160)(0,0)
\special{"
gnudict begin
gsave
50 50 translate
0.100 0.100 scale
0 setgray
/Helvetica findfont 100 scalefont setfont
newpath
-500.000000 -500.000000 translate
LTa
480 251 M
0 1858 V
LTb
480 251 M
63 0 V
2874 0 R
-63 0 V
480 483 M
63 0 V
2874 0 R
-63 0 V
480 716 M
63 0 V
2874 0 R
-63 0 V
480 948 M
63 0 V
2874 0 R
-63 0 V
480 1180 M
63 0 V
2874 0 R
-63 0 V
480 1412 M
63 0 V
2874 0 R
-63 0 V
480 1645 M
63 0 V
2874 0 R
-63 0 V
480 1877 M
63 0 V
2874 0 R
-63 0 V
480 2109 M
63 0 V
2874 0 R
-63 0 V
480 251 M
0 63 V
0 1795 R
0 -63 V
1067 251 M
0 63 V
0 1795 R
0 -63 V
1655 251 M
0 63 V
0 1795 R
0 -63 V
2242 251 M
0 63 V
0 1795 R
0 -63 V
2830 251 M
0 63 V
0 1795 R
0 -63 V
3417 251 M
0 63 V
0 1795 R
0 -63 V
480 251 M
2937 0 V
0 1858 V
-2937 0 V
480 251 L
LT0
3174 1946 D
597 1281 D
715 1433 D
832 1582 D
950 1539 D
1067 1509 D
1185 1421 D
1302 1349 D
1420 1353 D
1655 1263 D
1890 1233 D
2125 1342 D
2360 1278 D
2595 1176 D
2830 1039 D
3065 1052 D
3300 952 D
3114 1946 M
180 0 V
-180 31 R
0 -62 V
180 62 R
0 -62 V
597 1025 M
0 511 V
566 1025 M
62 0 V
-62 511 R
62 0 V
87 -309 R
0 412 V
684 1227 M
62 0 V
-62 412 R
62 0 V
86 -240 R
0 366 V
801 1399 M
62 0 V
-62 366 R
62 0 V
87 -377 R
0 302 V
919 1388 M
62 0 V
-62 302 R
62 0 V
86 -318 R
0 274 V
-31 -274 R
62 0 V
-62 274 R
62 0 V
87 -338 R
0 225 V
-31 -225 R
62 0 V
-62 225 R
62 0 V
86 -277 R
0 185 V
-31 -185 R
62 0 V
-62 185 R
62 0 V
87 -179 R
0 182 V
-31 -182 R
62 0 V
-62 182 R
62 0 V
204 -355 R
0 349 V
-31 -349 R
62 0 V
-62 349 R
62 0 V
204 -336 R
0 263 V
-31 -263 R
62 0 V
-62 263 R
62 0 V
204 -182 R
0 317 V
-31 -317 R
62 0 V
-62 317 R
62 0 V
204 -339 R
0 235 V
-31 -235 R
62 0 V
-62 235 R
62 0 V
204 -321 R
0 202 V
-31 -202 R
62 0 V
-62 202 R
62 0 V
2830 941 M
0 196 V
2799 941 M
62 0 V
-62 196 R
62 0 V
3065 979 M
0 147 V
3034 979 M
62 0 V
-62 147 R
62 0 V
3300 882 M
0 140 V
3269 882 M
62 0 V
-62 140 R
62 0 V
LT1
480 716 M
30 0 V
29 0 V
30 0 V
30 0 V
29 0 V
30 0 V
30 0 V
29 0 V
30 0 V
30 0 V
29 0 V
30 0 V
30 0 V
29 0 V
30 0 V
30 0 V
29 0 V
30 0 V
30 0 V
29 0 V
30 0 V
30 0 V
29 0 V
30 0 V
30 0 V
29 0 V
30 0 V
30 0 V
29 0 V
30 0 V
30 0 V
29 0 V
30 0 V
30 0 V
29 0 V
30 0 V
30 0 V
29 0 V
30 0 V
30 0 V
29 0 V
30 0 V
30 0 V
29 0 V
30 0 V
30 0 V
29 0 V
30 0 V
30 0 V
29 0 V
30 0 V
30 0 V
29 0 V
30 0 V
30 0 V
29 0 V
30 0 V
30 0 V
29 0 V
30 0 V
30 0 V
29 0 V
30 0 V
30 0 V
29 0 V
30 0 V
30 0 V
29 0 V
30 0 V
30 0 V
29 0 V
30 0 V
30 0 V
29 0 V
30 0 V
30 0 V
29 0 V
30 0 V
30 0 V
29 0 V
30 0 V
30 0 V
29 0 V
30 0 V
30 0 V
29 0 V
30 0 V
30 0 V
29 0 V
30 0 V
30 0 V
29 0 V
30 0 V
30 0 V
29 0 V
30 0 V
30 0 V
29 0 V
30 0 V
480 1645 M
30 0 V
29 0 V
30 0 V
30 0 V
29 0 V
30 0 V
30 0 V
29 0 V
30 0 V
30 0 V
29 0 V
30 0 V
30 0 V
29 0 V
30 0 V
30 0 V
29 0 V
30 0 V
30 0 V
29 0 V
30 0 V
30 0 V
29 0 V
30 0 V
30 0 V
29 0 V
30 0 V
30 0 V
29 0 V
30 0 V
30 0 V
29 0 V
30 0 V
30 0 V
29 0 V
30 0 V
30 0 V
29 0 V
30 0 V
30 0 V
29 0 V
30 0 V
30 0 V
29 0 V
30 0 V
30 0 V
29 0 V
30 0 V
30 0 V
29 0 V
30 0 V
30 0 V
29 0 V
30 0 V
30 0 V
29 0 V
30 0 V
30 0 V
29 0 V
30 0 V
30 0 V
29 0 V
30 0 V
30 0 V
29 0 V
30 0 V
30 0 V
29 0 V
30 0 V
30 0 V
29 0 V
30 0 V
30 0 V
29 0 V
30 0 V
30 0 V
29 0 V
30 0 V
30 0 V
29 0 V
30 0 V
30 0 V
29 0 V
30 0 V
30 0 V
29 0 V
30 0 V
30 0 V
29 0 V
30 0 V
30 0 V
29 0 V
30 0 V
30 0 V
29 0 V
30 0 V
30 0 V
29 0 V
30 0 V
stroke
grestore
end
showpage
}
\put(3054,1946){\makebox(0,0)[r]{$U_{double}/U$}}
\put(1948,51){\makebox(0,0){d [a]}}
\put(3417,151){\makebox(0,0){25}}
\put(2830,151){\makebox(0,0){20}}
\put(2242,151){\makebox(0,0){15}}
\put(1655,151){\makebox(0,0){10}}
\put(1067,151){\makebox(0,0){5}}
\put(480,151){\makebox(0,0){0}}
\put(420,2109){\makebox(0,0)[r]{5}}
\put(420,1645){\makebox(0,0)[r]{4}}
\put(420,1180){\makebox(0,0)[r]{3}}
\put(420,716){\makebox(0,0)[r]{2}}
\put(420,251){\makebox(0,0)[r]{1}}
\end{picture}}
\caption{\label{b240_nd}Ratio of total field energies for static double and single charge pairs as a function of distance $d$ at $\beta=2.4$ on a $32^3$ lattice.}
\end{figure}
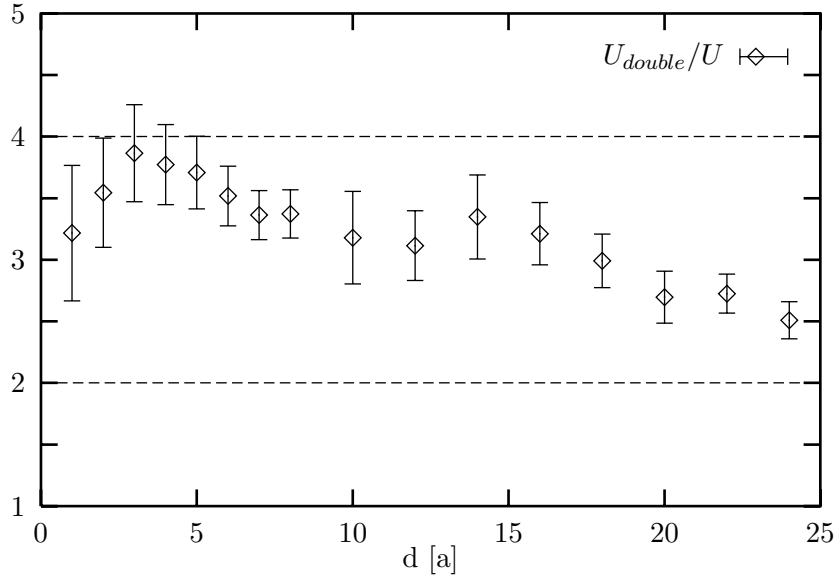

The Coulomb potential from static sources may further hide the proportionality of the string tension to the charge. Thus we have also computed the total field energy $U$ of a system with static sources on a $32^3$ lattice at $\beta=2.4$ and show the ratio $U_{double}/U$ as a function of the charge distance in fig.~\ref{b240_nd}. As demonstrated in the last section, we may expect that this ratio shows roughly the same behaviour as the one for the free energy. It can be seen that $U_{double}/U$ indeed is compatible with 4 (but not with 2) at smaller distances. For larger distances it takes values in an intermediate region around 3, which is in agreement with the behaviour of closed flux tubes at the given parameters shown in fig.~\ref{ratio_3d}.

Thus we have shown that the data obtained by our dual simulations quite agree with data from standard simulations, but within the dual formulation it is much easier to exclude systematic errors hiding the true nature of flux tubes.

\section{The interaction between flux tubes in four-dimensional $U(1)$}\label{ia}

For the three-dimensional theory our data have clearly demonstrated that the string tension scales proportionally to the charge, there is no detectable interaction between flux tubes. As already mentioned, the situation is different in four-dimensional $U(1)$. In figs.~\ref{ratio} and \ref{factor_u} we have seen a small deviation from the proportionality to the charge in the $\beta$-range slightly below the phase transition. We will now perform a closer investigation of the interaction between flux tubes for $\beta=0.96$. We will try to clarify the origin of this interaction, to calculate the interaction energy in the case of flux tubes between static double charges, and to investigate its influence on the flux tube profiles.

\begin{figure}
\centerline{\input{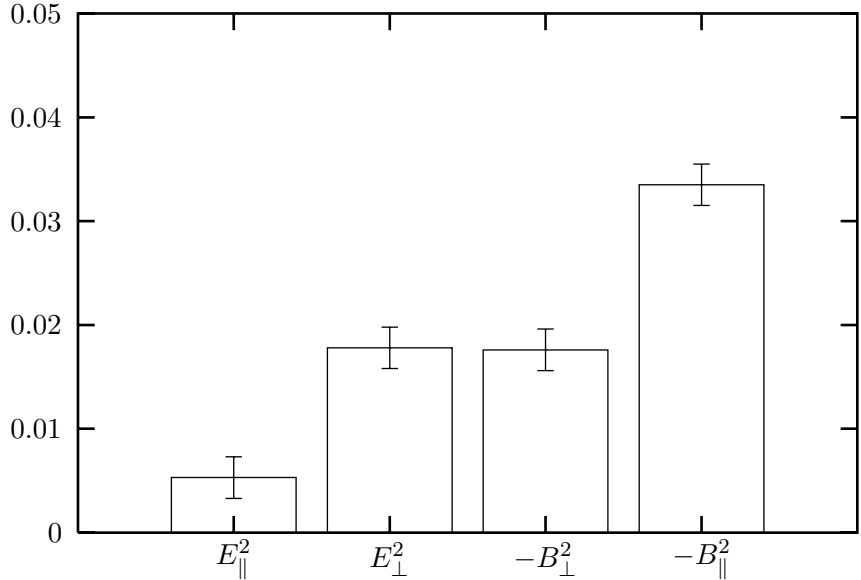}}
\caption{\label{ww_fsq}The individual contributions to the interaction energy $U_{double}-2U_{single}$ for a periodically closed double flux tube at $\beta=0.96$. The contribution of the electric field is positive, that of the magnetic field negative; therefore the total interaction energy is negative.}
\end{figure}

Let us first analyze the effect for periodically closed flux tubes. If we consider the total field energy, we are able to separate the interaction energy $U_{double}-2U_{single}$ into individual components -- electric and magnetic, longitudinal and transverse. The result is shown in fig.~\ref{ww_fsq}. There is almost no interaction term for the longitudinal electric field $E_\|^2$ which gives the main contribution to the string tension. The transverse components of electric and magnetic field cancel each other, due to the symmetry of the lattice. As can be seen the main contribution of the interaction between flux tubes is due to the longitudinal magnetic field $B_\|^2$. This increases the average action but decreases the average energy of a double flux tube. As discussed in ref.~\cite{pr97}, the longitudinal magnetic field only plays an important role at higher values of $\beta$ where the fluctuations in the dual model significantly increase. In terms of the dual simulation one can argue that these excitations of the string are enhanced by the presence of a second flux tube: It may happen that instead of 4 new plaquettes in a cube-like excitation only 2 plaquettes get occupied. It is plausible that such an opening of new channels also lowers the free energy of the system. The responsibility of $B_\|^2$ for the attractive interaction between flux tubes also makes clear why there is no such effect in three dimensions where we can only distinguish between three different types of fields ($E_\|$, $E_{\perp}$ and $B_{\perp}$).

\begin{figure}
\centerline{\input{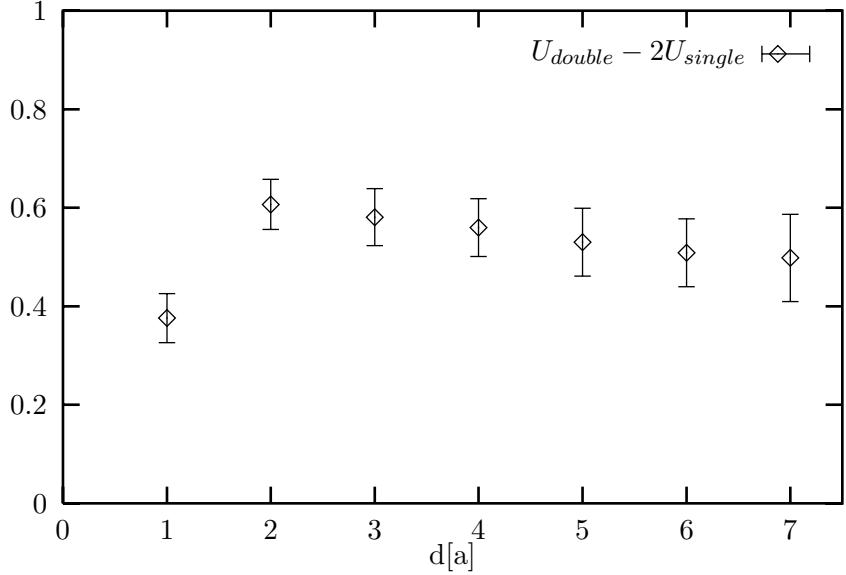}}
\caption{\label{ww_nd}The interaction (field) energy of a pair of double charges as a function of charge distance at $\beta=0.96$.} 
\end{figure}

If static double charges are attached to the ends of such a double flux tube, we expect an additional interaction term in the field energy, due to the repulsion between equal charges. The interaction energy $U_{double}-2U_{single}$ is shown as a function of the flux tube length $d$ in fig.~\ref{ww_nd}. A strong increase is observed till $d=2a$, where Coulombic behaviour is dominating, for larger distances the interaction energy seems to decrease slightly (although difficult to resolve even in the dual simulation, this effect is in agreement with the observations for closed flux tubes). Thus the positive interaction energy for a doubly charged system is only due to the self energy of the double charges, for very long flux tubes we might expect that this is compensated by the negative contribution of the flux tube attraction.

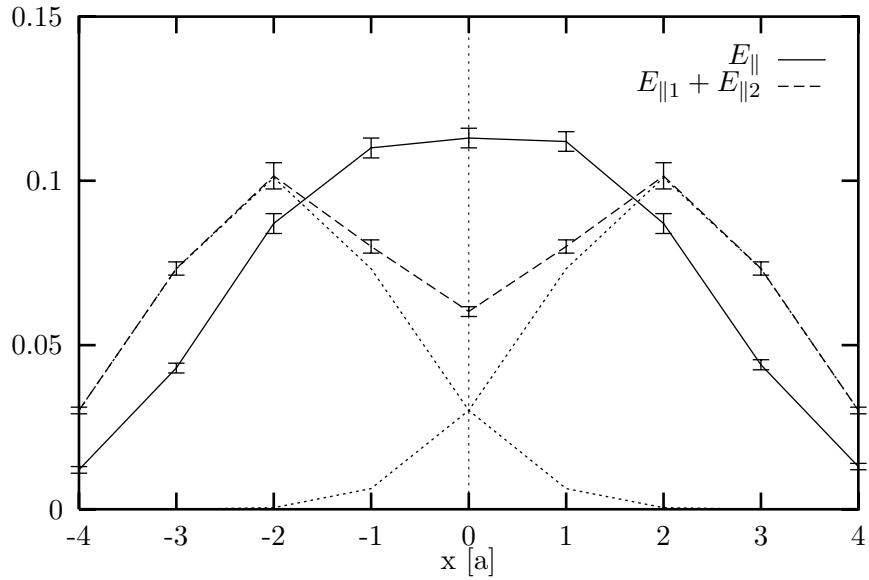
\begin{figure}
\centerline{
\setlength{\unitlength}{0.1bp}
\special{!
/gnudict 40 dict def
gnudict begin
/Color false def
/Solid false def
/gnulinewidth 5.000 def
/vshift -33 def
/dl {10 mul} def
/hpt 31.5 def
/vpt 31.5 def
/M {moveto} bind def
/L {lineto} bind def
/R {rmoveto} bind def
/V {rlineto} bind def
/vpt2 vpt 2 mul def
/hpt2 hpt 2 mul def
/Lshow { currentpoint stroke M
  0 vshift R show } def
/Rshow { currentpoint stroke M
  dup stringwidth pop neg vshift R show } def
/Cshow { currentpoint stroke M
  dup stringwidth pop -2 div vshift R show } def
/DL { Color {setrgbcolor Solid {pop []} if 0 setdash }
 {pop pop pop Solid {pop []} if 0 setdash} ifelse } def
/BL { stroke gnulinewidth 2 mul setlinewidth } def
/AL { stroke gnulinewidth 2 div setlinewidth } def
/PL { stroke gnulinewidth setlinewidth } def
/LTb { BL [] 0 0 0 DL } def
/LTa { AL [1 dl 2 dl] 0 setdash 0 0 0 setrgbcolor } def
/LT0 { PL [] 0 1 0 DL } def
/LT1 { PL [4 dl 2 dl] 0 0 1 DL } def
/LT2 { PL [2 dl 3 dl] 1 0 0 DL } def
/LT3 { PL [1 dl 1.5 dl] 1 0 1 DL } def
/LT4 { PL [5 dl 2 dl 1 dl 2 dl] 0 1 1 DL } def
/LT5 { PL [4 dl 3 dl 1 dl 3 dl] 1 1 0 DL } def
/LT6 { PL [2 dl 2 dl 2 dl 4 dl] 0 0 0 DL } def
/LT7 { PL [2 dl 2 dl 2 dl 2 dl 2 dl 4 dl] 1 0.3 0 DL } def
/LT8 { PL [2 dl 2 dl 2 dl 2 dl 2 dl 2 dl 2 dl 4 dl] 0.5 0.5 0.5 DL } def
/P { stroke [] 0 setdash
  currentlinewidth 2 div sub M
  0 currentlinewidth V stroke } def
/D { stroke [] 0 setdash 2 copy vpt add M
  hpt neg vpt neg V hpt vpt neg V
  hpt vpt V hpt neg vpt V closepath stroke
  P } def
/A { stroke [] 0 setdash vpt sub M 0 vpt2 V
  currentpoint stroke M
  hpt neg vpt neg R hpt2 0 V stroke
  } def
/B { stroke [] 0 setdash 2 copy exch hpt sub exch vpt add M
  0 vpt2 neg V hpt2 0 V 0 vpt2 V
  hpt2 neg 0 V closepath stroke
  P } def
/C { stroke [] 0 setdash exch hpt sub exch vpt add M
  hpt2 vpt2 neg V currentpoint stroke M
  hpt2 neg 0 R hpt2 vpt2 V stroke } def
/T { stroke [] 0 setdash 2 copy vpt 1.12 mul add M
  hpt neg vpt -1.62 mul V
  hpt 2 mul 0 V
  hpt neg vpt 1.62 mul V closepath stroke
  P  } def
/S { 2 copy A C} def
end
}
\begin{picture}(3600,2160)(0,0)
\special{"
gnudict begin
gsave
50 50 translate
0.100 0.100 scale
0 setgray
/Helvetica findfont 100 scalefont setfont
newpath
-500.000000 -500.000000 translate
LTa
480 251 M
2937 0 V
-1468 0 R
0 1858 V
LTb
480 251 M
63 0 V
2874 0 R
-63 0 V
480 870 M
63 0 V
2874 0 R
-63 0 V
480 1490 M
63 0 V
2874 0 R
-63 0 V
480 2109 M
63 0 V
2874 0 R
-63 0 V
480 251 M
0 63 V
0 1795 R
0 -63 V
847 251 M
0 63 V
0 1795 R
0 -63 V
1214 251 M
0 63 V
0 1795 R
0 -63 V
1581 251 M
0 63 V
0 1795 R
0 -63 V
1949 251 M
0 63 V
0 1795 R
0 -63 V
2316 251 M
0 63 V
0 1795 R
0 -63 V
2683 251 M
0 63 V
0 1795 R
0 -63 V
3050 251 M
0 63 V
0 1795 R
0 -63 V
3417 251 M
0 63 V
0 1795 R
0 -63 V
480 251 M
2937 0 V
0 1858 V
-2937 0 V
480 251 L
LT0
3114 1946 M
180 0 V
480 400 M
847 784 L
367 545 V
367 285 V
368 37 V
367 -13 V
367 -309 V
3050 796 L
3417 412 L
LT1
3114 1846 M
180 0 V
480 624 M
367 535 V
367 349 V
367 -266 V
1949 997 L
367 245 V
367 266 V
367 -349 V
3417 624 L
480 624 P
847 1159 P
1214 1508 P
1581 1242 P
1949 997 P
2316 1242 P
2683 1508 P
3050 1159 P
3417 624 P
480 611 M
0 25 V
449 611 M
62 0 V
-62 25 R
62 0 V
336 498 R
0 50 V
-31 -50 R
62 0 V
-62 50 R
62 0 V
336 275 R
0 99 V
-31 -99 R
62 0 V
-62 99 R
62 0 V
336 -341 R
0 50 V
-31 -50 R
62 0 V
-62 50 R
62 0 V
1949 978 M
0 37 V
-31 -37 R
62 0 V
-62 37 R
62 0 V
336 202 R
0 50 V
-31 -50 R
62 0 V
-62 50 R
62 0 V
336 192 R
0 99 V
-31 -99 R
62 0 V
-62 99 R
62 0 V
336 -424 R
0 50 V
-31 -50 R
62 0 V
-62 50 R
62 0 V
3417 611 M
0 25 V
-31 -25 R
62 0 V
-62 25 R
62 0 V
LT0
480 400 P
847 784 P
1214 1329 P
1581 1614 P
1949 1651 P
2316 1638 P
2683 1329 P
3050 796 P
3417 412 P
480 387 M
0 25 V
449 387 M
62 0 V
-62 25 R
62 0 V
847 765 M
0 37 V
816 765 M
62 0 V
-62 37 R
62 0 V
336 489 R
0 75 V
-31 -75 R
62 0 V
-62 75 R
62 0 V
336 210 R
0 75 V
-31 -75 R
62 0 V
-62 75 R
62 0 V
337 -37 R
0 74 V
-31 -74 R
62 0 V
-62 74 R
62 0 V
336 -87 R
0 74 V
-31 -74 R
62 0 V
-62 74 R
62 0 V
336 -384 R
0 75 V
-31 -75 R
62 0 V
-62 75 R
62 0 V
3050 777 M
0 38 V
-31 -38 R
62 0 V
-62 38 R
62 0 V
3417 400 M
0 24 V
-31 -24 R
62 0 V
-62 24 R
62 0 V
LT3
480 624 M
367 534 V
367 344 V
367 -344 V
1949 624 L
2316 329 L
367 -72 V
367 -6 V
367 0 V
480 251 M
367 0 V
367 6 V
367 72 V
368 295 V
367 534 V
367 344 V
367 -344 V
3417 624 L
stroke
grestore
end
showpage
}
\put(3054,1846){\makebox(0,0)[r]{$E_{\| 1}+E_{\| 2}$}}
\put(3054,1946){\makebox(0,0)[r]{$E_{\|}$}}
\put(1948,51){\makebox(0,0){x [a]}}
\put(3417,151){\makebox(0,0){4}}
\put(3050,151){\makebox(0,0){3}}
\put(2683,151){\makebox(0,0){2}}
\put(2316,151){\makebox(0,0){1}}
\put(1949,151){\makebox(0,0){0}}
\put(1581,151){\makebox(0,0){-1}}
\put(1214,151){\makebox(0,0){-2}}
\put(847,151){\makebox(0,0){-3}}
\put(480,151){\makebox(0,0){-4}}
\put(420,2109){\makebox(0,0)[r]{0.15}}
\put(420,1490){\makebox(0,0)[r]{0.1}}
\put(420,870){\makebox(0,0)[r]{0.05}}
\put(420,251){\makebox(0,0)[r]{0}}
\end{picture}}
\caption{\label{d4}Longitudinal electric field profile of two interacting flux tubes in the symmetry plane ($E_{\|}$, solid line). The length of flux tubes is $d=22a$, the transverse distance of equal charges is $4a$. For comparison, the dotted lines show the results for single flux tubes at $x=-2a$ and $x=+2a$, and the dashed line corresponds to the superposition $E_{\| 1}+E_{\| 2}$ of these two non-interacting flux tubes.}
\end{figure}

Finally we return to the investigation of flux tube profiles. Since two flux tubes were shown to attract each other, we should expect that two (initially) separated flux tubes prefer to form one single flux tube. To test this scenario we consider two charge pairs at a transverse distance of 4 lattice spacings in $x$-direction. It is important to investigate long flux tubes which have sufficient mobility to ``feel'' the interaction. Thus we take again $d=22a$ and measure the distribution of the longitudinal electric field along the $x$-axis in the symmetry plane. The result is depicted in fig.~\ref{d4}. It differs significantly from the field which is obtained by a superposition of the contributions of two single flux tubes, shifted in transverse direction by $\pm 2$ lattice spacings. This demonstrates that the system seems to prefer the formation of one flux tube, in analogy to a dual type-I superconductor.

\newpage

\section{Conclusions}

The dual formulation of $U(1)$ lattice gauge theory has proved to be a very efficient tool for the numerical investigation of doubly charged systems and the interaction between flux tubes. We are able to exclude strong finite temperature effects by considering Polyakov loops with large time extents; long flux tubes become numerically accessible; doubly charged systems can be studied with equal accuracy; a simple implementation of periodically closed flux tubes eliminates the contributions from static sources; finally we can calculate the total field energy and the free energy (potential) of flux tubes by dual simulations.

We have found that both in four and in three dimensions the string tension scales proportionally to the charge at strong coupling. Deviations from this behaviour in the energy of doubly charged systems in three-dimensional $U(1)$ have been shown to be due to finite size effects and to the influence of static sources. In four dimensions there is an attractive interaction between flux tubes, which becomes stronger towards the phase transition (a similar phenomenon has also been observed for $\mathbb{Z}_2$ gauge theory at criticality \cite{gliozzi}). This effect in $U(1)$ is due to the longitudinal component of the magnetic field; it can be ``hidden'' by the strong repulsion of the equal charges. For flux tubes of sufficient length we also demonstrated that a fusion into one flux tube is preferred.

In ref.~\cite{trottier} the result that the potential scales like the squared charge in three-dimensional $U(1)$ for the considered parameters was interpreted as a similarity between the confinement mechanism in Abelian and non-Abelian gauge theories. One should be very careful, however, in comparing the results. In $SU(3)$ lattice gauge theory it was shown that the Casimir scaling of the string tension can only be observed up to intermediate distances, where the flux tube cannot change its character \cite{mueller}. As the interquark distance increases, it becomes energetically favourable to pair-create gluons which screen higher representation sources. In addition, the phenomenon of flux tube coalescence, which is observed in non-Abelian gauge theories \cite{dfg}, is absent in $U(1)$.

Finally we want to comment on the interpretation of $U(1)$ lattice gauge theory in the confining phase as a dual superconductor. The numerical results we have presented show that flux tubes in four dimensions attract each other. This corresponds to the behaviour of a type-I superconductor. But this does not mean that $U(1)$ {\it is} a dual superconductor in the sense that it is described exactly by a dual Higgs model. It is well-known that the $U(1)$ partition function can be rewritten as a double limit (two couplings sent to infinity) of a dual Higgs model; as discussed in ref.~\cite{pr97} the corresponding transformation of expectation values shows that the agreement between the observables in the two models is not complete. Nevertheless, one can try to describe the observables of $U(1)$ lattice gauge theory approximately by an effective Higgs model with appropriately chosen parameters. According to our $U(1)$ results, we expect that such parameters should be those of a type-I superconductor.

\end{document}